\documentclass[apj, twocolappendix,numberedappendix]{emulateapj}

\usepackage{amsmath,amssymb}
\usepackage{xcolor}
\usepackage{mathptmx}
\usepackage{graphicx}
\usepackage[normalem]{ulem}

\usepackage[colorlinks,citecolor=blue]{hyperref}
\usepackage[all]{hypcap} 
\def\sectionautorefname~#1\null{Section #1\null}
\def\subsectionautorefname~#1\null{Section #1\null}
\def\subsubsectionautorefname~#1\null{Section #1\null}

\newcommand{\mum}{\ifmmode{\rm \mu m}\else{$\mu$m}\fi}

\newcommand{\sevenrm}{\rm\scriptsize}

\newcommand{\NeV}{[Ne{\sevenrm\,V}]}

\slugcomment{Submitted on Jan. 28, 2017; Accepted on Apr. 23, 2017; Published as \href{https://doi.org/10.3847/1538-4357/aa7051}{Lyu \& Rieke 2017, ApJ, 841, 76}}

\begin{document}

\title{
   The Intrinsic Far-infrared Continua of Type-1 Quasars
}

\author{Jianwei Lyu\altaffilmark{1} and George H. Rieke\altaffilmark{1}
}

\altaffiltext{1}{
    Steward Observatory, University of Arizona, 933 North Cherry Avenue,
    Tucson, AZ 85721, USA; \email{jianwei@email.arizona.edu}
}

\begin{abstract}
    The range of currently proposed active galactic nucleus (AGN) far-infrared
    templates results in uncertainties in retrieving host galaxy information
    from infrared observations and also undermines constraints on the outer
    part of the AGN torus.  We discuss how to test and reconcile these
    templates. Physically, the fraction of the intrinsic AGN IR-processed
    luminosity compared with that from the central engine should be consistent
    with the dust-covering factor. In addition, besides reproducing the
    composite spectral energy distributions (SEDs) of quasars, a correct AGN IR
    template combined with an accurate library of star-forming galaxy templates
    should be able to reproduce the IR properties of the host galaxies, such as
    the luminosity-dependent SED shapes and aromatic feature strengths. We
    develop tests based on these expected behaviors and find that the shape of
    the AGN intrinsic far-IR emission drops off rapidly starting at
    $\sim20~\mum$ and can be matched by an Elvis et al.-like template with a
    minor modification.  Despite the variations in the near- to mid-IR bands,
    AGNs in quasars and Seyfert galaxies have remarkably similar intrinsic
    far-IR SEDs at $\lambda \sim 20$-$100~\mum$, suggesting a similar emission
    character of the outermost region of the circumnuclear torus. The
    variations of the intrinsic AGN IR SEDs among the type-1 quasar population
    can be explained by the changing relative strengths of four major dust
    components with similar characteristic temperatures, and there is evidence
    for compact AGN-heated dusty structures at sub-kiloparsec scales in the
    far-IR.
\end{abstract}

\keywords
{galaxies:active --- infrared:galaxies --- quasars:general}

\section{Introduction}

\setcounter{footnote}{0}

\begin{figure}[htp]
    \begin{center}
	\includegraphics[width=1.00\hsize]{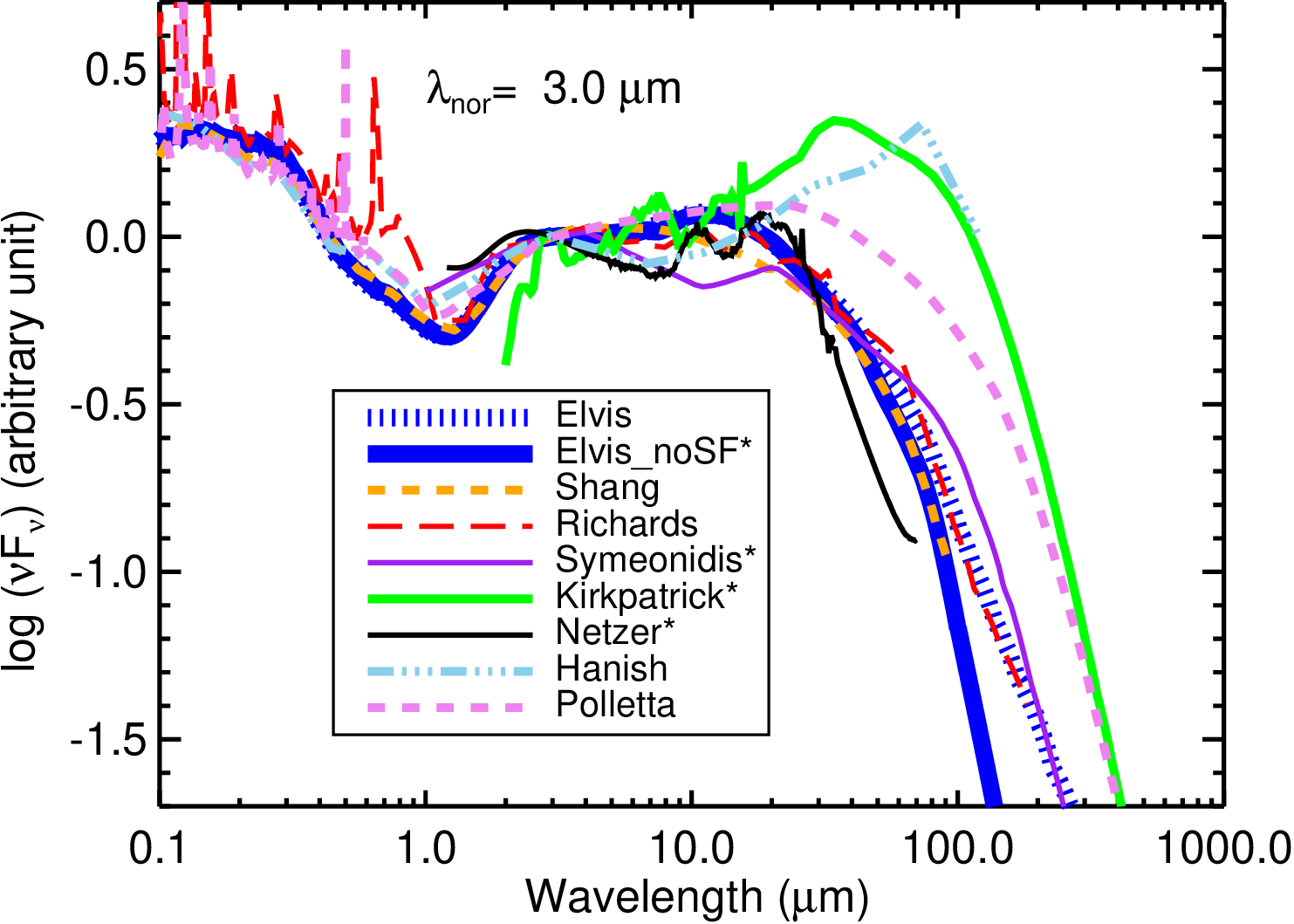} 
           \caption{
	    Examples of proposed empirical AGN SED templates, normalized at
	    3.0~$\mum$. The nine versions are \cite{Elvis1994} based partly on
	    the Palomar-Green (PG) sample; the same with a far-infrared
	    star-forming component removed (Elvis\_noSF) by \cite{Xu2015a};
	    \cite{Richards2006} based on 259 quasars; the radio-quiet quasar
	    template in \cite{Shang2011}; the star-formation-corrected AGN
	    template based on the radio-quiet PG sample proposed by
	    \cite{Symeonidis2016}; AGN3 from \cite{Kirkpatrick2015} with star
	    formation removed by subtracting the suitably normalized SFG2 from
	    the same source; \cite{Netzer2007} with star formation subtraction
	    in the far-IR; that proposed by \cite{Hanish2013}; and the QSO1
	    template from \cite{Polletta2007}. We plot those templates 
	    with efforts to remove the star-formation contamination as solid
	    lines and marked them with `*' in the legend.
	    }
	\label{fig:agn_temp_comp}
    \end{center}
\end{figure}

The infrared (IR) emission of quasars opens an invaluable window to study the
nature of the central active galactic nuclei (AGNs) as well as their host
galaxies. It is now widely accepted that the AGN is powered by gas accretion
onto the black hole and a substantial fraction of such accretion-released
energy is absorbed by the surrounding dusty structures and re-emitted in the
infrared (e.g., \citealt{Rieke1978}; \citealt{Neugebauer1986}). In the past
$\sim$ 30 years, with the launch and operation of each major space-based IR
telescope, many papers have appeared with the goal of characterizing the IR
spectral energy distributions (SEDs) of these systems, using data of improved
quality and/or larger samples (e.g., \citealt{Neugebauer1986, Sanders1989,
Elvis1994} with the {\it Infrared Astronomical Satellite} in 1980s; e.g.,
\citealt{Polletta2000, Haas2000, Haas2003, Alonso-Herrero2003} with the {\it
Infrared Space Observatory} in 1990s; e.g., \citealt{Richards2006,
Polletta2007, Netzer2007, Shang2011, Mullaney2011, Mor2012} with the {\it
Spitzer Space Telescope} in 2003-2009, e.g.,
\citealt{Petric2015,Garcia-Gonzalez2016} with the {\it Herschel Space
Observatory} in 2009-2013, and, e.g., \citealt{Mor2011, Petric2015} with the
{\it Wide-field Infrared Survey Explorer} in the post-2010). Nonetheless, these
intensive efforts have unexpectedly ended in significant disagreements about
the appropriate template, especially for the far-IR (e.g., from 20 to 1000
$\mu$m). This can be seen in Figure~\ref{fig:agn_temp_comp}, where we provide a
partial summary of the AGN empirical IR templates in the literature.

The possible different levels of contamination from the IR light emitted by the
host galaxies could underlie the broad distribution of the far-IR SEDs among
different templates. Several groups have attempted to remove this
contamination from the AGN template \citep[e.g.,][]{Netzer2007, Mullaney2011,
Kirkpatrick2015, Xu2015a, Symeonidis2016}. However, there is still no
consensus. These AGN ``intrinsic'' IR templates range from (1) the minimalist
far-infrared output derived by \cite{Netzer2007} by subtracting a
starburst-dominated ULIRG template from the average quasar IR SED under the
assumption that the vast majority of the 50--100~$\mum$ emission is due to star
formation, to (2) versions with substantially more far-infrared emission
obtained, for example, by \cite{Kirkpatrick2015} who fitted the data with a
combination of a galaxy mid-infrared spectrum, a power law in the same region
for the AGN, and two black bodies at longer wavelengths. Most recently, based on
a similar low-$z$ sample of optically selected bright quasars,
\cite{Symeonidis2016} reported an ``intrinsic'' AGN IR template with more
far-IR emission than that predicted by the classic quasar template from
\cite{Elvis1994}, where the latter did not even correct the contamination by
host galaxy star formation.

The importance of identifying the correct far-IR behavior of AGNs cannot be
overstated. As pointed out by \cite{Lutz2014} and \cite{Netzer2015},
theoretical torus models have too much freedom to distribute dust in ways that
may not reflect the real situations. In fact, given the wide range of
observational results on AGN IR SEDs, no convincing observational tests can be
made for these model templates in the far-IR.  Determining which type of far-IR
behavior is appropriate (or whether there is a broad range) will therefore have
a direct impact on models of AGN tori, especially on their relation to the
surrounding interstellar material.  To measure the IR star formation rates
(SFRs) of quasar host galaxies, a proper handling of the AGN far-IR
contribution is critical. A number of papers have analyzed the quasar IR SEDs
with a combination of the SED library from some torus model and
templates/models for the host galaxies, and derived the host SFRs from the IR
luminosities of the latter component \citep[e.g.,][]{Delvecchio2014,
Leipski2014, Dong2016, Netzer2016}.  However, since none of the torus models
have been observationally well constrained in the far-IR, the accuracy of the
SFR measurements is not always guaranteed, even if the fitting residuals are
tiny. Recently, there have been concerns regarding whether the AGN contributes
substantially to the heating of the surrounding interstellar medium (ISM;
\citealt{Schneider2015,Roebuck2016,Symeonidis2017}). The validity of such
studies also rests on the knowledge of the complete SED of the intrinsic AGN IR
emission.

As shown by our previous work \citep{Lyu2017}, it is clear that one single AGN
IR template does not apply to all the type-1 quasar population ($L_{\rm
AGN}\gtrsim10^{11}L_\odot$). Additionally, similar SED variations of the
intrinsic AGN IR emission are seen among quasars from $z\sim$0 to $z\sim$6,
suggesting that such behavior is common. Following that work, this paper
examines the intrinsic far-IR SEDs of AGNs in detail with the focus mainly on
the type-1 quasar population. Firstly, an energy balance argument is applied to
test some proposed AGN SED templates in Section~\ref{sec:energy}.  Then we
present an in-depth analysis of the far-IR emission of the Palomar--Green (PG)
quasars based on the study of \cite{Lyu2017}, clarify several issues regarding
the derivation of a correct intrinsic AGN far-IR SED, and determine our
preferred template in Section~\ref{sec:decomp}.  In Section~\ref{sec:discuss},
we discuss whether the AGN far-IR intrinsic template based on quasars is
applicable to other kinds of AGNs, as well as some implications for the
obscuring structure from our results.  A summary is given in
Section~\ref{sec:summary}. We provide the intrinsic templates for type-1
quasars in Appendix~\ref{app:temp}.

\section{Energy balance}\label{sec:energy}

\subsection{Rationale}

Luminous AGNs are believed to be powered by gas accretion onto the central
supermassive black holes \citep{Salpeter1964, Lynden-Bell1969, Shields1978}.
Various processes associated directly with such central engines dominate the
X-ray to the optical far-red energy output of these systems. The continuum
emission of the central engine should drop substantially at
$\lambda\gtrsim1$~$\mu$m as suggested from multiple theoretical studies
\citep[e.g.,][]{Kawaguchi2001, Cao2009,Liu2012} and confirmed by observations
\citep{Kishimoto2008}.  Under the precepts of AGN unification (e.g.,
\citealt{Antonucci1993}; see the recent review by \citealt{Netzer2015}), the
central engine is surrounded by equatorial optically thick dusty structures
that cause the nuclear photons to escape preferentially along the polar
direction. Meanwhile, the energy of those ``blocked" photons is absorbed by
dust and reradiated into the near infrared to 10 $\mu$m (and almost certainly
beyond to at least $\sim$ 100 $\mu$m) as a result of energy balance.

For a type-1 AGN viewed face-on, the SED crossover between the direct
(intrinsic) emission from the accretion disk and the IR-processed emission by
the dust obscuration is close to the spectral minimum at $\sim$ 1.3~$\mum$.
This basic hypothesis for the source emission was first suggested by SEDs
\citep{Rieke1978}, then by detailed decomposition into spectral components
demonstrating a reduction in the amplitude of the variability as one goes
further into the infrared \citep{Rieke1981, Cutri1985, Neugebauer1989,
Neugebauer1999}. It has been firmly established through reverberation mapping,
which shows time lags in the infrared response to variations in the
optical--ultraviolet output by roughly the time expected if the reradiating dust
is at the distance from the nucleus where it is heated close to its sublimation
temperature \citep[e.g.,][]{Koshida2014, Gandhi2015, Lira2015, Pozo2015,
Schnulle2015}.

If the infrared emission of quasars is reradiated energy absorbed from the
central engine and accretion disk, a corollary is that the infrared luminosity
must be less than that of the central source, or more accurately, less than the
energy absorbed from the central source. The geometry of the torus is
often estimated by comparing the energy emitted by the central engine at
wavelengths $<$1.3~$\mu$m with that emitted in the infrared, presuming that the
central engine emits isotropically and that the infrared emission is dominated
by energy absorbed in the torus.  This approach is supported not only by the
rapidly declining emission from the accretion disk going from the optical into
the near-IR, but also by the low levels of absorption by silicate-rich
interstellar dust in the red and near-IR \citep[e.g.,][]{Corrales2016}.
Therefore, it is the optical blue, UV, and X-ray that are most effective in
providing the energy input to the dust that obscures the nucleus. The portion
of the total accretion disk emission that is reprocessed into the infrared is
represented by the luminosity of the system between 1.3 and 1000 um, as seen in
a type-1 AGN, that is $f_{\rm R}=L_{\rm dust}/L_{\rm accr.disk}$. When the dust
is optically thick, $f_{\rm R}$ can be interpreted as the covering factor for
the torus, $f_c$. As discussed in Section~\ref{sec:bal-result}, a more accurate
way to interpret $f_{\rm R}$ needs to take account of possible anisotropies in
the accretion disk radiation and of radiative transfer in the torus
\citep[e.g.,][]{Runnoe2013, Stalevski2016}.

As will be shown below, a number of recently proposed AGN SED templates with
strong far-IR emission fail the requirements of energy balance between the
accretion disk and the infrared luminosity.

\subsection{Template for AGN Soft X-Ray/UV/Optical Emission}

Since \cite{Elvis1994} published their SED template for unobscured type-1
(broad-line) quasars, there have been multiple studies that have confirmed the
uncanny accuracy of this work \citep[e.g.,][]{Richards2006, Shang2011,
Runnoe2012, Hanish2013, Scott2014, Lyu2017}. As a result, we can take the
intrinsic SED of a type-1 quasar to be well-understood from the soft X-ray to
$\sim$ 10 $\mu$m. \cite{Lusso2013} show that the torus is probably optically
thin to X-rays more energetic than $\sim$ 1 keV, so they should not be included
in this energy budget argument. Consequently, we can avoid concerns that the
Elvis template may be biased toward being X-ray bright \citep{Marconi2004}. 

The properties of AGNs show possible trends in SED behaviors from the
most luminous quasars to the much less luminous low-ionization nuclear
emission-line region (LINERs) galaxies \citep{Ho1999, Ho2008}. However, the
\cite{Elvis1994} template has been developed on quasars with $L_{\rm
AGN}\gtrsim10^{11}~L_\odot$ and all the following templates are also based on
AGN samples with similar luminosity ranges. Thus, we adopt the
\cite{Elvis1994} template to represent the accretion disk emission 
where some other template does not cover the soft X-ray to optical bands.

\subsection{Energy Budget Calculations}
\label{sec:bal-exam}

Table~\ref{tab:balance} summarizes the sample properties, approaches to 
synthesize template SEDs, and calculations of the energy budgets for the proposed AGN templates
listed below.  Nearly every template needs some adjustment or extension for
these energy budget determinations. Because the luminosity is a free parameter, we have
normalized the results so the luminosity between 1.3 $\mu$m and 1 keV is set to
1. As a result, the tabulated infrared luminosities indicate directly the
reradiated fraction of the emission of the central engine.

\capstartfalse
\begin{deluxetable*}{p{50pt}p{80pt}p{90pt}p{10pt}p{70pt}p{20pt}p{80pt}p{30pt}}
\tabletypesize{\scriptsize}
    \tablecaption{\label{tab:balance} Fraction of Short-$\lambda$ Luminosity Reradiated in the Infrared of Representative AGN Templates}
\tablehead{
    Template  &  $\lambda$ Range    &  Sample Characteristics  &  $N$   &   Stack Method   &  SF Cor.?   & Correction Base & $f_{\rm R}$  \\
    (1)       & (2)            &  (3)                & (4)  &  (5)             &  (6)                 &   (7)             &   (8)
}
\startdata
    Elvis                         &  X-ray to Radio   & optical-blue, X-ray bright quasars         &    29   & Kaplan--Meier Mean, $\lambda_{\rm nor}=1.25~\mum$      &  N   & --  &  $\ge$0.55  \\
    Elvis\_noSF &  X-ray to Radio   & optical-blue, X-ray bright quasars         &    29   & Kaplan-Meier Mean, $\lambda_{\rm nor}=1.25~\mum$      &  Y   & relation between 11.3 aromatic features and the IR 50-25~$\mum$ flux ratio of both quasars and SFGs   &   $\ge$0.53 \\
    Richards                      &  X-ray to Radio   & mid-IR and optical color-selected quasars  &  259    & Gap repair, luminosity matched    &  N   & --  & $\ge$0.65   \\
    Shang                         &  X-ray to Radio   & UV/optical bright quasars                  &   27    & Median Mean, $\lambda_{\rm nor}=0.42~\mum$   &  N   & --  & $\ge$0.64    \\
    Symeonidis                    &  0.4-500~$\mum$   & radio-quiet PG quasars                     &   47    & Arithmetic Mean, $\lambda_{\rm nor}=20~\mum$ &  Y   & Matching the 11.3 aromatic feature-derived SFR with the SFG templates IR luminosity  &   $\ge$0.56    \\
    Kirkpatrick AGN1              &   2-1000~$\mum$   & $z\sim$0.80 (U)LIRGs w/ AGN dominated (100\%) mid-IR spectra & 22  & Median Mean, normalized by the 5-15~$\mum$ luminosity & Y  & Decomposition of the mid-IR spectra & $\ge$0.70-0.78  \\
    Kirkpatrick AGN2              &   2-1000~$\mum$   & $z\sim$1.03 (U)LIRGs w/ AGN dominated (93\%) mid-IR spectra & 23  & Median Mean, normalized by the 5-15~$\mum$ luminosity & Y  & Decomposition of the mid-IR spectra & $\ge$1.12-1.02  \\
    Kirkpatrick AGN3              &  2-1000~$\mum$    & $z\sim$1.65 (U)LIRGs w/ AGN dominated (94\%) mid-IR spectra &  21    & Median Mean, normalized by the 5-15~$\mum$ luminosity & Y  & Decomposition of the mid-IR spectra & $\ge$0.80-0.85  \\
    Kirkpatrick AGN4              &  2-1000~$\mum$    & $z\sim$1.96 (U)LIRGs w/ AGN dominated (93\%) mid-IR spectra & 31  & Median Mean, normalized by the 5-15~$\mum$ luminosity &  Y  & Decomposition of the mid-IR spectra & $\ge$1.34-1.26  \\
    Netzer                        &  1.2-70~$\mum$    &  PG quasars, missing very high-luminosity objects &   29 &   Arithmetic Mean, $\lambda_{\rm nor}=6~\mum$ &  Y  & Assuming the 50-100~$\mum$ emission of the IR-weak and IR-strong quasar template is due to star formation & $\ge$0.46   \\
    Hanish                        &  0.1-100~$\mum$  &  SDSS spectroscopically selected quasars without strong optical reddening  & 301  & Median Mean, normalized by the 0.2-1.0~$\mum$ luminosity & N  & --  &  $\ge$0.91   \\
    Polletta QSO1                 & X-ray to mid-IR   & Optically selected, spectroscopically confirmed type-1 quasars from SDSS                        &   35    & Mean weighted by luminosity, $\lambda_{\rm nor}=1.0~\mum$ &  N  &  --  &$\ge$0.74 
\enddata
    \tablecomments{ Column (4): the size of the sample from which the template was
	derived; Column (8): IR-processed light fraction of the template.
	$f_{\rm R}=L_{\rm dust}/L_\text{accr. disk}$. We calculate the 1
	keV--1.25~$\mum$ luminosity of the AGN template as $L_\text{accr. disk}$
	and the 1.25--1000~$\mum$ luminosity as $L_\text{dust}$. For the Kirkpatrick templates, 
          we show the results for normalizing to the Elvis template between 2 and 10 $\mu$m and then at 12 $\mu$m.}
\end{deluxetable*}
\capstarttrue

{\bf \cite{Elvis1994}}: the procedure is most straightforward in the case of
the \cite{Elvis1994} template. Table~\ref{tab:balance} shows two results, the
first for the original form and the second for the star-formation-removed
version from \cite{Xu2015a}.

{\bf \cite{Richards2006}}: it was necessary to fill out this template for
frequencies log($\nu$) above 17 and below 12.5; this was done by joining on
sections of the original Elvis template. 

{\bf \cite{Shang2011}}: this work updated the \cite{Elvis1994} SED with data
from HST, {\it FUSE}, {\it Spitzer}, {\it Chandra}, and {\it XMM}. The silicate
emission features in the mid-IR were reproduced thanks to the spectral data
from {\it Spitzer}/IRS. Although the \cite{Elvis1994} SED was based on IRAS
far-IR data with many upper limits, these two SED templates agreed remarkably
well in the far-IR. There is no need to adjust this template.

{\bf \cite{Symeonidis2016}}: this template was completed for wavelengths short
of 0.4 $\mu$m by joining on an original Elvis template. The far-infrared
behavior is similar to that of the Polletta template, but the lower fraction of
reradiated luminosity results because the template is relatively bright in the
0.3--1.3 $\mu$m range (see Figure~\ref{fig:agn_temp_comp}).

{\bf \cite{Kirkpatrick2015}}: these templates were built according to the dominance
of the AGN continuum contribution in mid-IR spectra from a study of 343
(ultra)luminous infrared galaxies at $z$=0.3--2.8, without constraints on the
AGN types. Since the Kirkpatrick templates are only given for 2--1000 $\mu$m, to
estimate the strength of the intrinsic disk accretion emission, we need to
combine them with the Elvis template at shorter wavelengths with proper
scalings. To do so, we forced the total luminosity between 2 and 10 $\mu$m to
be the same, and then added the Elvis template luminosity between 1.3 and 2
$\mu$m to the reradiated total for the \cite{Kirkpatrick2015} template. 
Their sample probably includes some fraction of obscured or type-2 AGNs.
Under the unified theory of AGNs, we expect these objects to have identical
intrinsic X-ray/UV/optical and infrared reradiated SEDs to type-1 objects. 
However, if they have strong mid-infrared extinction, 
the normalization at 2--10 $\mu$m could underestimate the nuclear emission.
Nevertheless, considering the lack of strong silicate absorption features in
the Kirkpatrick AGN templates, such extinction is unlikely to be significant
\citep[e.g.,][]{Shi2006, Hatziminaoglou2015}.  In addition, the rough agreement
of the continuum in these templates with the Elvis one between 2 and 10~$\mum$
supports the normalization we have adopted to obtain an estimate of the
X-ray/UV/optical luminoisty. Finally, it is believed that the Kirkpatrick
templates may have a significant contamination from star formation in the host
galaxy (see the AGN template in their Figure 13). We therefore subtracted their
star forming template SFG2 until the long wavelength cutoff of the remaining
SED fell similarly to the Elvis template, i.e., we attributed as much as
possible of the far-infrared flux to star formation (see
Figure~\ref{fig:agn_temp_comp} for an illustration). The maximal nature of this
adjustment is clear since it made the resulting AGN-only template go negative
at wavelengths longer than 400 $\mu$m.  The two other star forming templates in
\cite{Kirkpatrick2015} are very similar in the far-infrared and using one of
them would not have yielded any significant differences.

{\bf \cite{Netzer2007}}:  since this template does not remove host galaxy
emission in the near-IR, we replace it with the Elvis template at
$\lambda<3.0~\mum$, where no spectral features exist and the host galaxy
contamination is negligible. We also believe the SED rise starting at
$\lambda\sim50~\mum$ is non-physical, so we completed the template with a
Rayleigh--Jeans tail with a wavelength-dependent emissivity proportional to
$\lambda^{-1.5}$ at $\lambda>40~\mum$.

{\bf \cite{Hanish2013}}: the template was completed for wavelengths shorter than
0.1 $\mu$m and longer than 120 $\mu$m by joining on segments of the original
Elvis template. The possibility of the far-infrared peak in the template
arising through star formation is mentioned by the authors, but there has been
no attempt to correct for it. It is therefore not surprising that there might
be a significant far-infrared component not associated with the central engine.

{\bf \cite{Polletta2007}}: this template was completed for wavelengths short of
0.1 $\mu$m with the Elvis template. There is no explicit step in the assembly
of the Polletta templates to remove a star-formation-powered  component in the
far-infrared, so it would not be surprising if there is a significant one.

\subsection{Analysis and Results}\label{sec:bal-result}

Assuming the standard AGN unification point of view that type-2 AGNs are the
same as type-1 AGNs but obscured by dust, a dust-covering factor can be
obtained from the relative fraction of type-2 (obscured) to type-1 (unobscured)
objects.  \cite{Schmitt2001} found that 70\%  of their far-IR-selected sample
of 88 Seyfert galaxies are obscured.  Since this sample is IR-selected, this
value is probably biased high.  Based on the optical spectra of SDSS galaxies,
\cite{Hao2005b} show that Seyfert 1 and Seyfert 2 galaxies have comparable
numbers at low luminosity, while Seyfert 1 galaxies outnumber Seyfert 2
galaxies by a factor of two to four at high luminosity.  \cite{Reyes2008} reported a
type-2 quasar fraction in the SDSS sample within the range of $\sim$ 0.5--0.6.
Considering the decreasing dust-covering factor with increasing AGN luminosity
as first noted by \cite{Lawrence1991} and confirmed by later works (e.g.,
\citealt{Maiolino2007}), plus the discussion in \citet{Stalevski2016}, we adopt
0.65 as an upper limit for the dust-covering factor ($f_{\rm c}$) in luminous AGNs. 

Comparison of the value of the dust-covering factor with the ratio of the
AGN infrared to optical--UV--X-ray luminosities is not straightforward because of
anisotropies in the radiation field of the central engine.
\citet{Stalevski2016} computed a grid of SEDs emitted by the dusty structures
and studied how the relation between the covering fraction and the reradiated
luminosity changes with different parameter values for the torus.  When the
accretion disk and the obscuring torus both emit isotropically, the
IR-processed light fraction is a perfect proxy for the dust-covering factor
with a one-to-one relation between these two quantities. In the more realistic
case of anisotropic accretion disk emission and a torus that is optically thick
in the mid-IR, their simulations suggest that these two quantities are in
agreement at $\sim$0.65. They also provide polynomial fits to allow estimation
of the expected values away from 0.65. Although somewhat bright in the
mid-infrared and faint in the far-infrared, the \citet{Stalevski2016} model
SEDs still match reasonably well relative to the \citet{Elvis1994}-like
observational AGN template, suggesting their results are appropriate for our
study. We have compared with the model results for $\tau_{9.7}$ of 3--10
since the resulting SEDs roughly match the templates.  The corresponding upper
limit to the ratio of reradiated to central engine luminosity (a.k.a. $f_{\rm
R}$) is then $\sim$ 0.75.

The templates from \cite{Elvis1994}, \cite{Symeonidis2016}, and
\cite{Netzer2007} satisfy the upper limit for energy balance between the output
of the central engine and the luminosity reradiated in the infrared. It appears
that the quasar templates by \cite{Hanish2013} contain too much far-IR
emission, possibly due to the contamination from the host galaxy star
formation. The \citet{Polletta2007} template is also only marginally consistent
with this constraint.  Although the exact values of the IR-processed light
fractions of the \cite{Kirkpatrick2015} templates have relatively large
uncertainties, they are still more luminous in the infrared than expected.
Consequently, from the perspective of energy balance, we suggest the
\cite{Kirkpatrick2015} templates are unlikely to represent the intrinsic far-IR
emission for common type-1 AGNs. One way to make these latter three template
families more consistent with energy balance is if their soft X-ray emission is
significantly stronger than indicated by the AGN templates. The energy
balance argument is difficult to apply to other AGN types because their
intrinsic X-ray/UV SEDs are hard to determine. In the case of the
\cite{Kirkpatrick2015} templates, we would underestimate the available X-ray/UV
luminosity only if there are enough obscured AGNs in the sample to suppress the
mid-IR spectrum where we have normalized the Elvis template.

To reassure ourselves of the credibility of the energy balance arguments
above, we have examined the behavior of NGC 4151, since (1) its UV continuum
has been derived in detail \citep{Alexander1999}, (2) its flux variations have
been exceptionally well monitored \citep[e.g.,][]{Lyuty1999,Doroshenko2001},
and (3) its infrared SED is very well determined \citep[e.g.,][]{Rieke1981,
Mcalary1983, Deo2009, Garcia-Gonzalez2016}. The energy balance test on
this archetypal type-1 AGN seems to fail at very first glance: even if we
take the UV continuum toward the upper range as derived by
\citet{Alexander1999}, the infrared luminosity is too large to be consistent
with energy balance by 20\% or more if we analyze the comparison for the simple
case of isotropic emission. However, this conclusion is fallacious.
Considering the diverse physical scales of the panchromatic emission and
the time variability of the accretion disk emission, energy balance for a
given AGN may not hold at a given time; the infrared output, particularly at
the longer wavelengths, represents a time average of the input UV luminosity.
If we normalize the \citet{Alexander1999} continuum to the average {\it
U}-band brightness from 1968 through 2000 \citep{Lyuty1999,Doroshenko2001} - a
factor of 1.62 higher than at the time of their study - ratios of $\sim$ 85\%
can be obtained but still with a UV continuum toward the upper range allowed
\citep{Alexander1999}. This value is consistent with the escape fraction of
15.7\% derived by \citet{Alonso-Herrero2011}. Rather than raising the UV
to the maximum allowed, another solution would be to increase the soft
X-ray flux. However, since most of the energy is produced in the blue and
UV, a substantial boost (by a factor of $\sim$3) would be necessary to make a
significant difference in the energy balance. Besides the variability
correction, we also need to consider the anisotropic nature of the accretion
disk emission. If we analyze the results as in the models of
\citet{Stalevski2016}, the values become consistent with more probable fits to
the UV continuum \citep{Alexander1999} and without increasing the soft
X-rays. As illustrated by this example, it is critical to apply the
energy balance test on a time-averaged SED. (In fact, this requirement is
satisfied by the construction of the AGN template.)  The case of NGC 4151 also
shows that the models by \citet{Stalevski2016} are appropriate to yield
consistent energy balance results.

Finally, our emphasis has been the intrinsic AGN SED: the spectrum emitted by
the central engine and circumnuclear torus that together appear to constitute a
typical active nucleus. It is possible that an AGN heats the surrounding ISM,
producing an additional emission component \citep[e.g.,][]{Roebuck2016}.
Because the extent of this heating will depend on parameters such as the
relative orientation of the circumnuclear torus and the host galaxy and on the
amount of interstellar material in the host, the emission will differ
significantly from one AGN to another and is likely to be insignificant in many
cases. We therefore do not consider it to be part of the intrinsic SED,
although it is an interesting phenomenon that should yield additional insights
to AGN behavior.

\section{Decomposition of the Quasar IR SEDs}\label{sec:decomp}

We illustrate some issues in deriving intrinsic AGN IR SEDs by extending the
discussion in \cite{Lyu2017} on the infrared SEDs of PG quasars.

\subsection{The $f_\text{\rm AGN, MIR}$--$f_\text{\rm AGN, TIR}$ Relation}

The mid-IR spectral window preserves a lot of useful information about the
star-forming activities in AGN host galaxies. With the aid of mid-IR
spectral decompositions or measurements of the aromatic feaures, many
authors have tried to remove the host galaxy contamination in the AGN far-IR
emission \citep[e.g.,][]{Mullaney2011, Kirkpatrick2012, Kirkpatrick2015,
Xu2015a, Symeonidis2016}.

\begin{figure}[htp]
   \begin{center}
	\includegraphics[width=1.0\hsize]{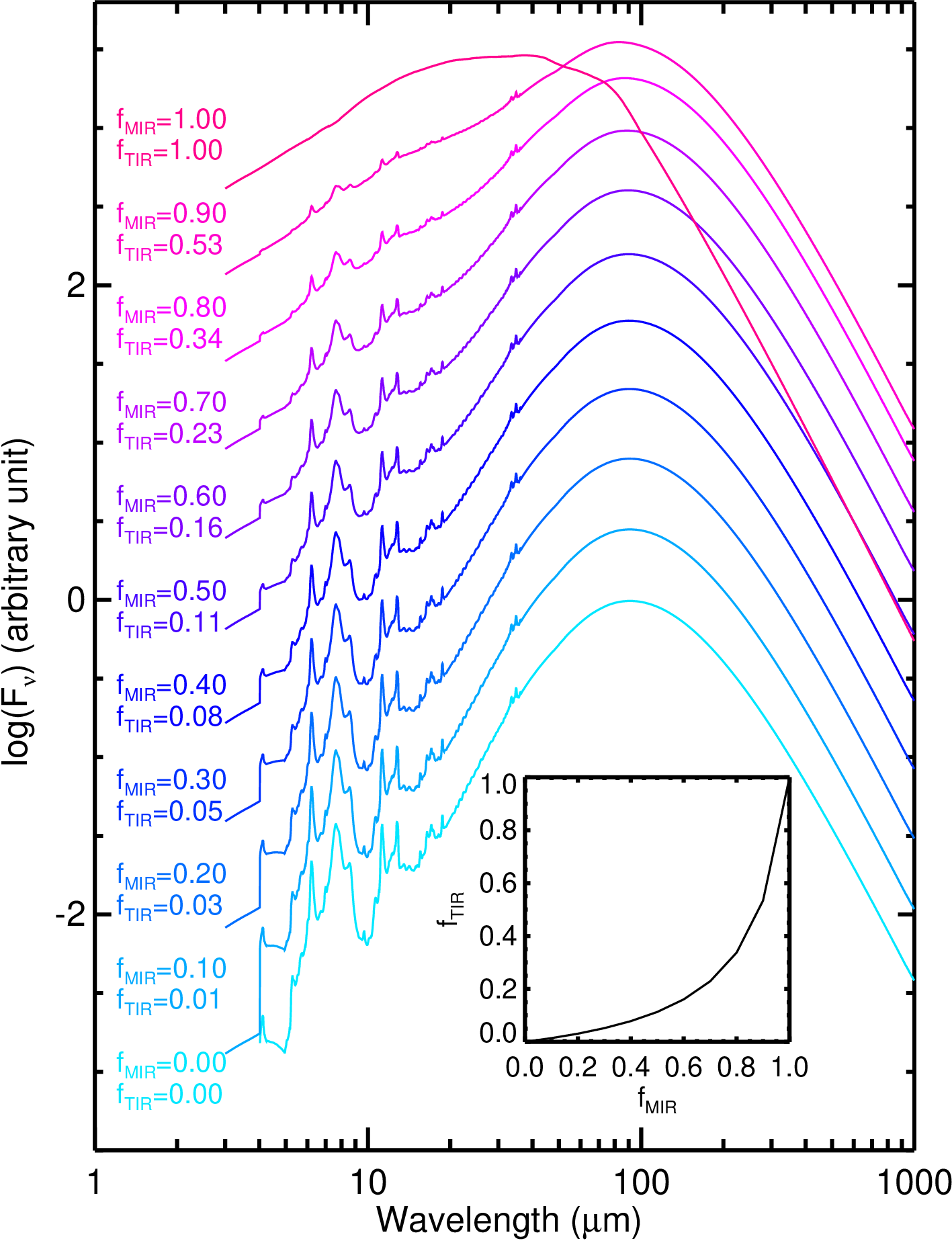}
	\caption{
	    Mock infrared SEDs of galaxies with different mixing of the Elvis
	    AGN template and the $\log (L_{\rm IR}/L_\odot) = 11.25$
	    \cite{Rieke2009} template. The AGN contributions in the mid-IR
	    ($f_{\rm MIR}$) and total-IR ($f_{\rm TIR}$) are denoted on the left
	    side of each SED.  The zoom-in panel shows the relation between
	    $f({\rm AGN})_{\rm MIR}$ and $f({\rm AGN})_{\rm TIR}$ in the mock
	    SEDs. 
	}
	\label{fig:agn_sf_f}
   \end{center}
\end{figure}

However, as suggested in Section 3.1 of \cite{Lyu2017}, the host contribution
in the infrared 8--1000$~\mum$ range could be underestimated by merely focusing
on the mid-IR spectral features. This argument is based on the different
IR SED behaviors of the AGN and the host galaxy: the IR output of an AGN is
peaked in the mid-IR but drops quickly in the far-IR, while the galaxy emission
is relatively weak in the mid-IR but strong in the far-IR. We illustrate this
issue in Figure~\ref{fig:agn_sf_f}, which shows a series of mock galaxy SEDs
derived by changing the relative contributions of the SF-corrected
\cite{Elvis1994} template \citep{Xu2015a} and a $\log (L_{\rm IR}/L_\odot) =
11.25$ \cite{Rieke2009} star-forming template.  While the continua shapes of
the 5--10$~\mum$ SEDs and the relative strengths of the aromatic features change
drastically with the AGN contribution in the mid-IR, there is little
modification in the far-IR.  The zoom-in panel shows how the relative AGN
contribution in the 8--1000~$\mum$, $f_{\rm TIR}$, changes with that in the
mid-IR (5--40$~\mum$), $f_{\rm MIR}$.  The host galaxy still contributes
$\sim$50\% of the 8--1000~$\mum$ luminosity of the composite SED when the AGN
provides 90\% of the mid-IR emission. This result qualitatively matches the
trend between the AGN relative contribution in the mid-IR emission and that in
the total IR emission of the infrared luminous galaxies, as observationally
determined by \citet[Section 5]{Kirkpatrick2015}.

For removing the host galaxy far-IR contribution based on relatively low S/N
mid-IR spectra, such non-linear relations between $f_\text{\rm AGN, MIR}$ and
$f_\text{\rm AGN, TIR}$ could cause an overestimation of the intrinsic AGN
emission. In other words, although the mid-IR star formation feature may not be
detected due to the low quality of the mid-IR data, substantial far-IR host
galaxy contamination is still possible. This effect could be one of the
reasons behind the strong far-IR emission in the \cite{Kirkpatrick2015}
AGN templates. In addition, since their derivation was based on LIRGs and
ULIRGs, there may be a selection bias toward cases where the AGN may be heating
the galactic ISM, as suggested by \cite{Kirkpatrick2015} and
\cite{Roebuck2016}. Nonetheless, this situation is also subject to the energy
balance constraint, which we have found to be an issue for these templates. In
comparison, the study by \cite{Mullaney2011} simultaneously fitted the mid-IR
spectra and the far-IR photometry of their X-ray selected sample, resulting in
a far-IR AGN SED shape similar to the SF-corrected \cite{Elvis1994} template
(see Section~\ref{sec:other-agn}).

\subsection{Using the Mid-IR Aromatic Features to Trace the SF
Contribution to the Quasar Far-IR Emission}

The 11.3$~\mum$ aromatic feature strength seems to be the only SFR-related
spectral feature that is not strongly contaminated or influenced by the AGN
\citep[e.g.,][]{Diamond-Stanic2010, Esquej2014,Alonso-Herrero2014}. By matching
the observed aromatic strength with a library of star-forming galaxy (SFG)
templates, \cite{Shi2007, Shi2014} have constrained the host galaxy IR
luminosities in quasars and these results have been used to derive the intrinsic AGN
IR templates \citep{Xu2015a, Symeonidis2016}.

With the mid-IR spectrum of good quality as an essential prerequisite,
the key to correctly gauge the galaxy contribution to the quasar far-IR
emission is the luminosity conversion factor between the aromatic features and
the galaxy far-IR emission, which is dependent on the selection of the SFG
template. Additionally, there is a possibility that some of the feature
excitation is provided by the AGN. These two issues are addressed in this
section. Finally, we note that the measurements of the aromatic features can
have significant biases due to the method to quantify the mid-IR
dust emission continuum: typically, the spectral decomposition approach (e.g.,
{\it PAHFIT}) with a model of several different dust components yields a factor
of $\sim2$ larger value of the 11.3~$\mum$ aromatic flux compared to the
``interpolation continuum'' approach of fitting some smooth function (e.g.,
spline or power law) to anchor points without strong aromatic emission
\citep[e.g.,][]{Smith2007}. We will utilize the results from spectral
decomposition similar to {\it PAHFIT} in \cite{Shi2014}, except in
Section~\ref{sec:sym-temp}.

\subsubsection{Choosing the Proper Star-Forming Galaxy Templates}\label{sec:SFG-temp}

Many IR templates for SFGs are available in the literature (e.g.,
\citealt{Chary2001, Dale2002, Lagache2003, Lagache2004, Siebenmorgen2007,
Rieke2009}; see a review in \citealt{Casey2014}, as well as
\citealt{Ciesla2014}).  A critical issue is that the 11.3 $\mu$m feature lies
on the edge of the silicate absorption. If this absorption is strong, the
apparent equivalent width (EW) of the 11.3~$\mum$ feature may be underestimated.
This problem is particularly acute in deconvolving a composite AGN and star
forming galaxy SED, where the apparent depth of the silicate absorption is
masked by the AGN emission. It becomes increasingly important at high IR
luminosities (e.g., LIRG/ULIRG; \citealt{Stierwalt2013}). Not all of the SFG
template libraries include silicate absorption (e.g., \citealt{Chary2001,
Dale2002}; see the discussion in Appendix 1.3 of \citealt{Rieke2009}).
Consequently, if SFG templates that omit the effect of silicate
absorptions were used, the far-IR emission from the galaxy can be
underestimated by adopting the wrong conversion factors from the aromatic
feature flux.

\begin{figure}[htp]
    \begin{center}
	\includegraphics[width=1.0\hsize]{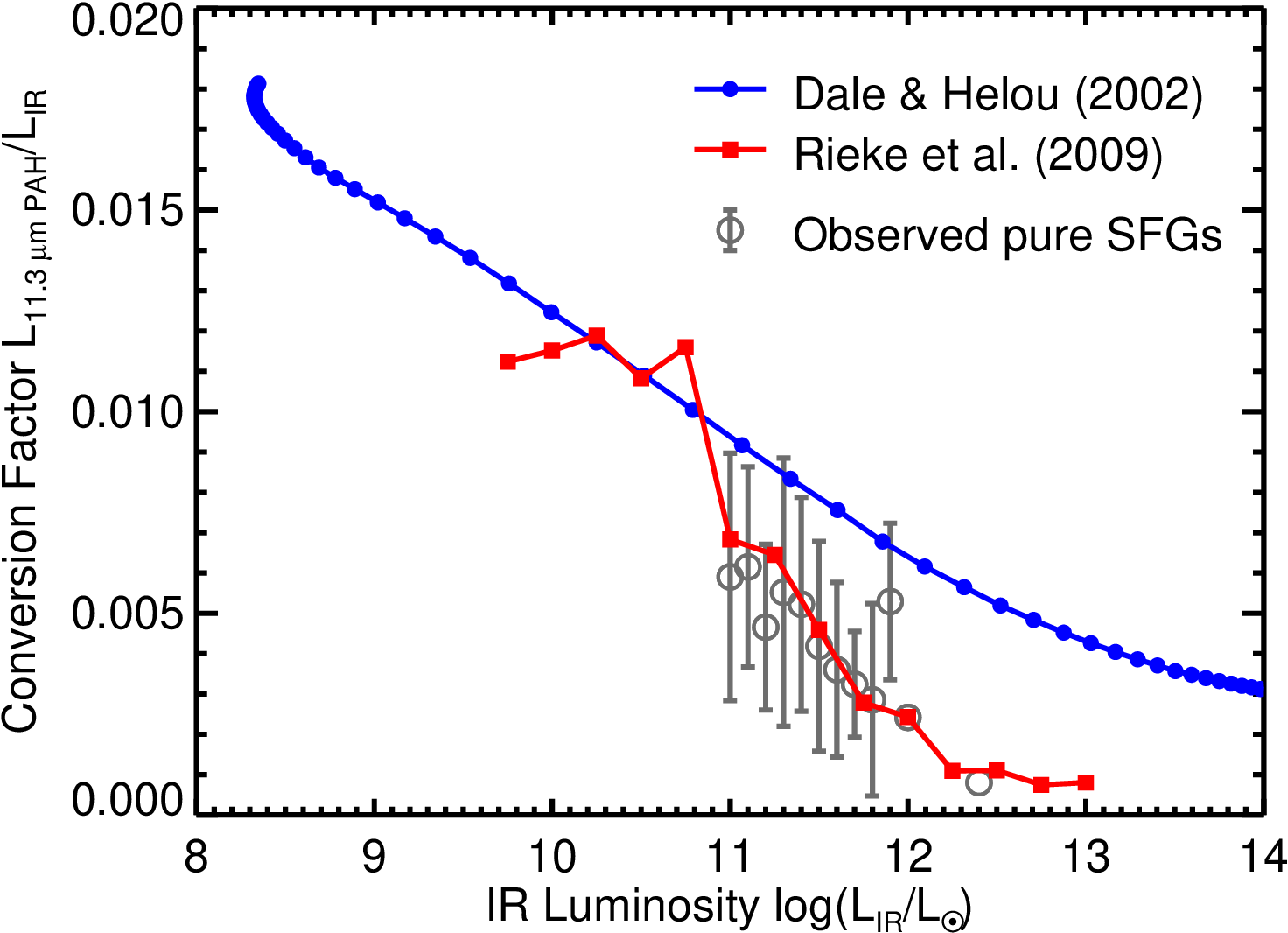}
          	\caption{
	    Comparisons of the luminosity conversion factors for the
	    11.3~$\mum$ aromatic feature to the 8--1000~$\mum$ IR emission in
	    \cite{Dale2002} templates and \cite{Rieke2009} templates.  We
	    also plot the observed correlation based on a study of $\sim$100
	    pure star-forming galaxies in the GOALS sample (see
	    Appendix~\ref{app:goals}). 
	}
	\label{fig:SFG_temp_comp}
    \end{center}
\end{figure}

With a consistent treatment of IR spectra, photometry, and theoretical models,
\cite{Rieke2009} developed IR templates for local galaxies based on {\it
Spitzer} data that include the accompanying increasing silicate absorption with
increasing IR luminosity. The accuracy of the \cite{Rieke2009} templates has
been demonstrated in a number of works \citep[e.g.,][]{Willmer2009,
Calzetti2010} and their validity to represent SFGs extends to redshifts up to 3
\citep[e.g.,][]{Rex2010,Rujopakarn2013, Sklias2014, Lyu2016}. In
Appendix~\ref{app:goals}, we derive the conversion factors for $\sim100$ pure
SFGs in the Great Observatories All-sky LIRG Survey (GOALS;
\citealt{Armus2009}). As shown in Figure~\ref{fig:SFG_temp_comp}, it is very
clear that the \cite{Dale2002} templates do not match the
observations.\footnote{We note that the \cite{Dale2002} templates were not
designed for use at the highest luminosities in the figure.} In contrast, the
\cite{Rieke2009} templates are in good agreement with the observed trend just
as expected. Compared with \cite{Rieke2009}, the \cite{Dale2002} SFG templates
would result in significantly lower estimates of the SFRs and
resulting far-infrared luminosities for galaxies with star forming luminosities
of $\sim$ 10$^{11}$ $L_\odot$ and higher.

For quasars, the discrepancy discussed here arises only if the host galaxy IR
star forming luminosities are 10$^{11}$ $L_\odot$ or more. The SFRs of
most PG quasar host galaxies were found to be $\sim$10--100~$M_\odot{\rm
yr}^{-1}$ (see \citealt{Shi2014}), putting them into the (U)LIRG category.
Moreover, studies that assume Elvis-like intrinsic AGN SEDs and that the
far-IR luminosity is dominated by star formation also suggest that many host
galaxies are in this high-luminosity range \citep[e.g.,][]{Xu2015a,
Lyu2016}.

The significance of the effects on the derivation of the intrinsic AGN far-IR
SED due to the selection of SFG templates will be demonstrated in
Section~\ref{sec:sym-temp}. 

\subsubsection{The Luminosity Dependence of the Host Galaxy IR SEDs}

According to the energy balance discussed in Section~\ref{sec:energy}, it is
not reasonable to assume the entire luminosities of far-IR-bright quasars are
attributed to the central engine. A significant, perhaps dominant, part of
the far-IR luminosity must arise from the host galaxy. To confirm that the
\cite{Rieke2009} SFG templates can represent the quasar host galaxy IR emission
and the aromatic emission observed in the quasar mid-IR spectra is indeed from
the quasar host galaxy, in a statistical sense, (1) the fitted host galaxy IR
luminosities should be consistent with the IR luminosities associated with the
selected SFG templates from \cite{Rieke2009}; (2) the 11.3~$\mum$
aromatic feature strengths observed in the mid-IR spectra should also agree
with the values converted from the fitted galaxy template. 

In \cite{Lyu2017}, the mid- to far-IR SEDs of 87 $z<0.5$ PG
quasars were fitted with the SF-corrected Elvis AGN template (as well as the
dust-deficient AGN templates) and SFG templates with $\log_{10}(L_{\rm
IR}/L_\odot)$=9.75--12.0 from \cite{Rieke2009}. Besides the well-sampled IR
broadband SEDs, high-quality {\it Spitzer}/IRS mid-IR spectra are also
available for the whole sample, enabling simultaneous tests on the
ability of both the far-IR SEDs and the mid-IR spectral features of the
\cite{Rieke2009} templates to represent the quasar host galaxy emission.
However, not all members in this sample were suitable for this purpose.
Firstly, for each quasar, we should be able to reveal the differences among the
\cite{Rieke2009} templates. Thus, we require the quasar to either have a
substantial host galaxy contribution ($f_\text{\rm IR, host}>33\%$), where the
shape of the infrared galaxy template becomes important to change the $\chi^2$,
or the \cite{Rieke2009} template with a similar luminosity as the observed
value tends to a smaller $\chi^2$ at least by a factor of 1.5 compared with
alternatives. 32 PG quasars meet this requirement. Additionally, there should
be no nearby galaxy to pollute the {\it Herschel} photometry of the quasar
far-IR emission, and two quasars failing this criteria have been dropped.
Finally, poor fittings of the very far-IR SEDs may suggest dust heating by old
stars, abnormal dust properties, confusion noise, or radio synchrotron emission
contamination.  In such cases, the radio-quiet AGN template combined with any
selection of the SFG templates will be unreliable.  Consequently, we dropped
another six quasars due to their large fitting residuals ($\gtrsim0.3$ dex) at
$\lambda>100~\mum$.  The final sample of the 24 PG quasars as well as their IR
properties is listed in Table~\ref{tab:PG-host-IR}. 

\capstartfalse
\begin{deluxetable*}{clcccccccc}
    \tabletypesize{\scriptsize}
    \tablewidth{1.0\hsize}
    \tablecolumns{10}
    \tablecaption{Host Galaxy IR Properties of the 24 Palomar--Green Quasars\label{tab:PG-host-IR}
    }
    \tablehead{
	\colhead{ID} &
	\colhead{Source} & 
	\colhead{$z$}  & 
	\colhead{$F_{\rm PAH,spec}$} & 
	\colhead{$L_{\rm R09, PAH}$} 	& 
	\colhead{$c_{\rm R09, PAH}$}    &
	\colhead{$f_{\rm IR, AGN}$} &
	\colhead{$F_{\rm PAH,SED}$} &
	\colhead{$L_{\rm R09, temp}$} 	& 
	\colhead{$L_{\rm host, obs.}$} \\
	\colhead{(1)} & 
	\colhead{(2)} & 
	\colhead{(3)} & 
	\colhead{(4)} & 
	\colhead{(5)} & 
	\colhead{(6)} &
	\colhead{(7)} &
	\colhead{(8)} &
	\colhead{(9)} &
	\colhead{(10)} 
    }
    \startdata
   4 &  PG 0043+039  &   0.38 &  $0.10 \pm 0.04 $ & 11.50 & 1.02  &  0.65  & 0.19 & 11.00&  11.56     \\
   6 &  PG 0050+124  &   0.06 &  $4.60 \pm 0.18 $ & 11.25 & 1.09  &  0.56  & 8.90 & 11.00&  11.47     \\
  10 &  PG 0838+770  &   0.13 &  $0.25 \pm 0.05 $ & 10.50 & 0.87  &  0.51  & 0.77 & 11.00&  11.12     \\
  14 &  PG 0923+129  &   0.03 &  $1.82 \pm 0.06 $ &  9.75 & 1.15  &  0.46  & 4.34 & 10.50&  10.33     \\
  15 &  PG 0934+013  &   0.05 &  $0.58 \pm 0.04 $ &  9.75 & 1.12  &  0.38  & 1.10 & 11.00&  10.39     \\
  20 &PG 1011$-$040  &   0.06 &  $0.56 \pm 0.04 $ & 10.00 & 0.88  &  0.53  & 0.86 & 11.00&  10.45     \\
  21 &  PG 1012+008  &   0.19 &  $<0.05         $ & 10.00 & 0.92  &  0.76  & 0.23 & 11.00&  10.96      \\
  22 &  PG 1022+519  &   0.05 &  $1.18 \pm 0.03 $ & 10.00 & 1.03  &  0.35  & 0.87 & 11.00&  10.29     \\
  25 &PG 1049$-$005  &   0.36 &  $<0.04         $ & 11.00 & 0.83  &  0.69  & 0.23 & 11.75&  11.98    \\
  31 &  PG 1119+120  &   0.05 &  $0.72 \pm 0.10 $ & 10.00 & 0.82  &  0.60  & 1.60 & 11.00&  10.56     \\
  34 &PG 1149$-$110  &   0.05 &  $0.14 \pm 0.04 $ &  9.75 & 0.26  &  0.44  & 1.32 & 11.00&  10.47     \\
  41 &  PG 1244+026  &   0.05 &  $0.52 \pm 0.04 $ &  9.75 & 0.91  &  0.58  & 0.69 & 11.00&  10.19     \\
  48 &  PG 1341+258  &   0.09 &  $0.21 \pm 0.04 $ &  9.75 & 1.26  &  0.65  & 0.28 & 11.00&  10.34     \\
  49 &  PG 1351+236  &   0.05 &  $2.02 \pm 0.03 $ & 10.50 & 1.10  &  0.21  & 2.21 & 11.00&  10.70     \\
  50 &  PG 1351+640  &   0.09 &  $1.77 \pm 0.09 $ & 11.25 & 0.91  &  0.65  & 2.06 & 11.00&  11.20     \\
  53 &  PG 1402+261  &   0.16 &  $0.31 \pm 0.11 $ & 10.75 & 0.98  &  0.76  & 0.41 & 11.50&  11.21     \\
  54 &  PG 1404+226  &   0.10 &  $0.27 \pm 0.03 $ & 10.00 & 1.28  &  0.67  & 0.37 & 10.75&  10.32     \\
  56 &  PG 1415+451  &   0.11 &  $0.78 \pm 0.03 $ & 10.75 & 1.09  &  0.66  & 0.46 & 11.00&  10.74     \\
  67 &  PG 1519+226  &   0.14 &  $0.20 \pm 0.04 $ & 10.25 & 1.10  &  0.85  & 0.30 & 10.75&  10.55     \\
  70 &  PG 1543+489  &   0.40 &  $<0.02         $ & 10.50 & 0.87  &  0.55  & 0.35 & 11.75&  12.27     \\
  73 &  PG 1612+261  &   0.13 &  $0.35 \pm 0.03 $ & 10.50 & 1.19  &  0.60  & 0.77 & 11.25&  11.14     \\
  77 &  PG 1700+518  &   0.28 &  $1.41 \pm 0.17 $ & 13.00 & 0.88  &  0.74  & 0.38 & 11.75&  11.95     \\
  78 &  PG 1704+608  &   0.37 &  $0.09 \pm 0.06 $ & 11.25 & 1.09  &  0.81  & 0.12 & 12.00&  11.78     \\
  81 &  PG 2209+184  &   0.07 &  $0.58 \pm 0.04 $ & 10.00 & 1.33  &  0.57  & 0.58 & 09.75&  10.20     
\enddata                                    
    \tablecomments{
	Column (1): the object ID (see Table 2 in \citealt{Lyu2017} for the
	list of the whole PG sample); Column (2): object name; Column (3):
	redshift; Column (4): the 11.3~$\mum$ aromatic feature flux (unit:
	$10^{-13}$erg s$^{-1}$ cm$^{-2}$) as measured by \cite{Shi2014}; Column
	(5): the IR luminosity of the best-matched \cite{Rieke2009} template
	based on the 11.3~$\mum$ aromatic feature luminosity; Column (6): the
	scaling factor of the best-matched \cite{Rieke2009} template based on
	the 11.3~$\mum$ aromatic feature luminosity; Column (7): the AGN
	fractional contribution to the IR emission of each quasar; Column (8):
	the 11.3~$\mum$ aromatic feature flux (unit: $10^{-13}$erg s$^{-1}$
	cm$^{-2}$) derived from the SED modeling; Column (9): the IR luminosity
	of the best-fitted \cite{Rieke2009} template from the SED modeling;
	Column (10): the observed IR luminosity of the host galaxy from the SED
	modeling.
    }
\end{deluxetable*}
\capstarttrue

\begin{figure}[htp]
    \begin{center} 
	\includegraphics[width=1.0\hsize]{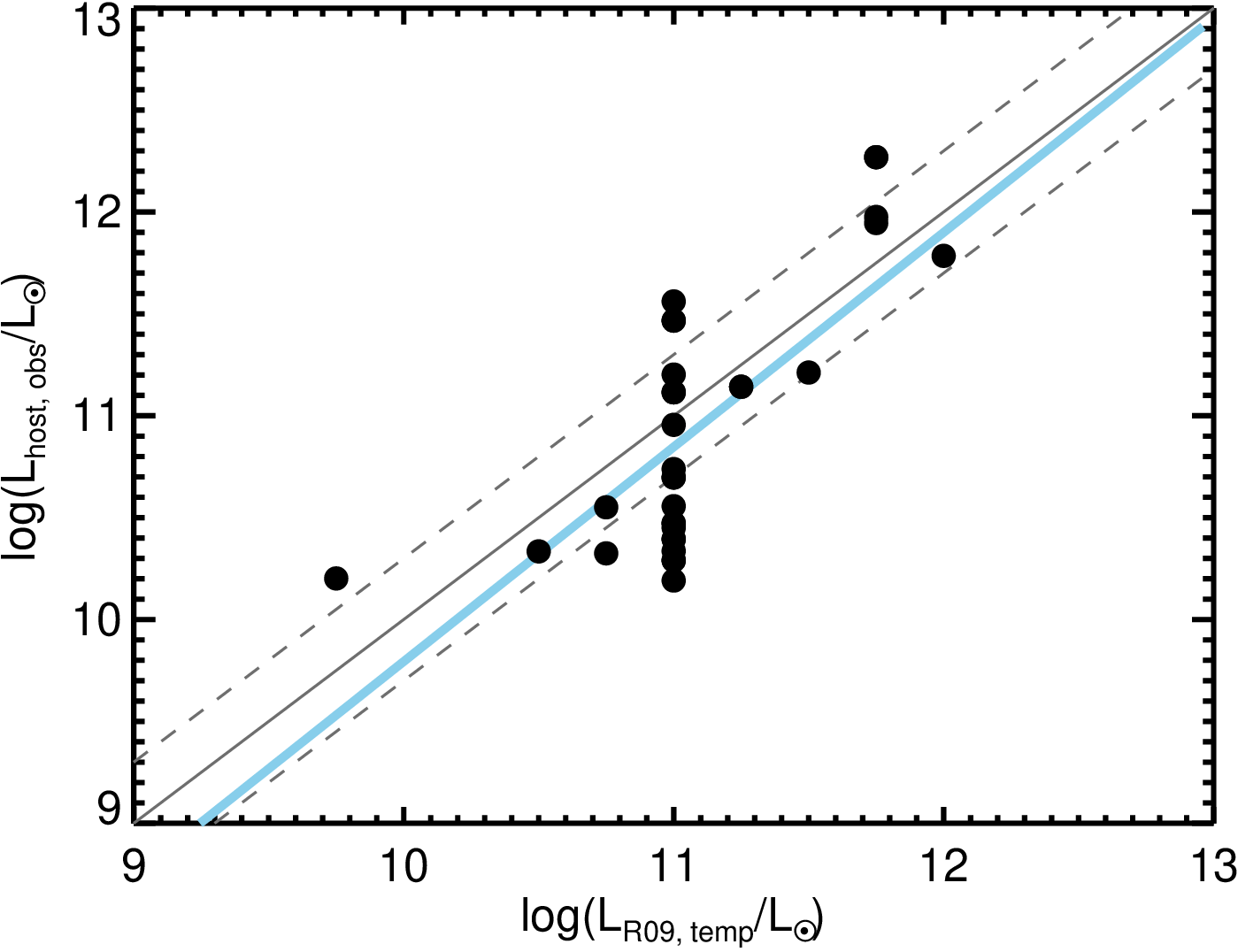} 
           \caption{ 
	    Relation between the derived host galaxy infrared luminosities
	    with the $\chi^2$ selected \cite{Rieke2009} template luminosities.
	    We show the 1:1 relation and a linear fitting of the data points as
	    gray and blue solid lines separately.
	    }
	\label{fig:ir_temp_select}
    \end{center} 
\end{figure}

The far-infrared SEDs of star-forming galaxies show a consistent pattern of
shapes as a function of the luminosity
\citep[e.g.,][]{Chary2001,Dale2002,Siebenmorgen2007,Rieke2009}. In
Figure~\ref{fig:ir_temp_select}, we compare the derived host galaxy infrared
luminosities from fitting the far-infrared SEDs with the luminosity of the
template with the shape that gave the best fit for these 24 PG quasars.  The
luminosities of the optimally shaped templates are roughly consistent with the
quasar host galaxy luminosities derived by integrating the fitted fluxes.  With
a linear fit, we find a slope between these two groups of luminosities to be
1.05$\pm$0.20 and an intercept of 0.73$\pm$2.13, consistent with the expected
1:1 relation.

\subsubsection{The 11.3~$\mum$ Aromatic Strength from the SED Model}

If the excitation of the 11.3~$\mum$ feature is dominated by star formation as
represented by the \cite{Rieke2009} templates, we should expect the observed
mid-IR aromatic feature strength to be consistent with the aromatic flux
converted from the SFG template fitted to the SED photometry. If the AGN plays
an important role in exciting the aromatic feature, we would expect the feature
strength to be greater than that predicted by the star-forming template.

To derive the aromatic feature strengths of the \cite{Rieke2009} templates, we
decomposed the 5--40~$\mum$ portion of the templates using the IDL program {\it
PAHFIT} \citep{Smith2007}. In this code, the mid-IR dust continuum is fitted
with multiple blackbody components with possible mid-IR extinction and the
11.3~$\mum$ aromatic feature is fitted with two Drude profiles centered at
11.23 and 11.33~$\mum$. The 11.3~$\mum$ feature strength computed by {\it
PAHFIT} was then compared with the 8.0--1000~$\mum$ integrated luminosity of
each template.  Finally, we ended up with the conversion factor for the
11.3~$\mum$ feature, varying from 0.011 at $\log_{10}(L_\text{\rm
IR,SF}/L_\odot)=9.75$ to 0.0024 at $\log_{10}(L_\text{\rm
IR,SF}/L_\odot)=12.0$.

\begin{figure}[htp]
    \begin{center}
	\includegraphics[width=1.0\hsize]{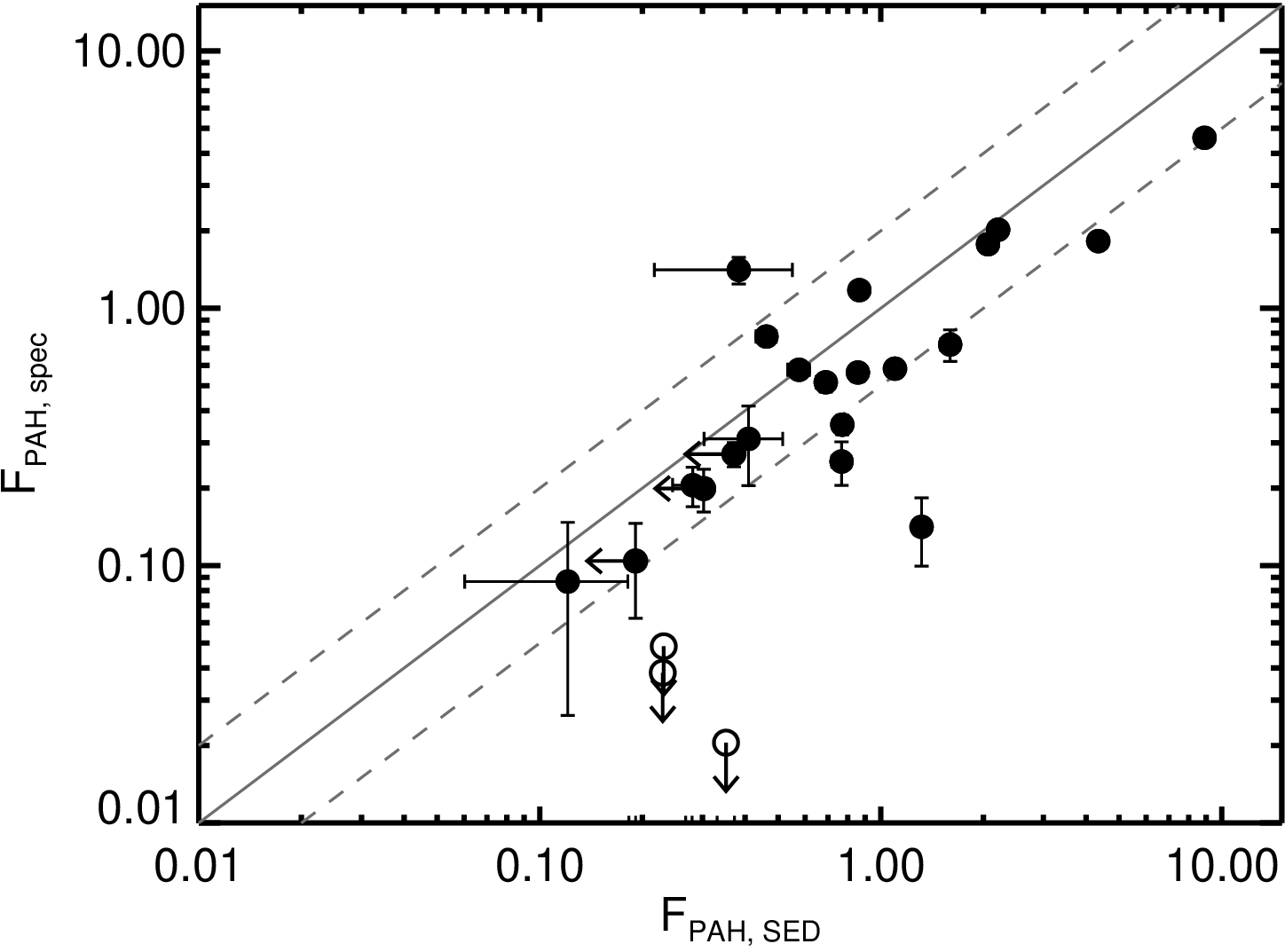}
	\caption{
	    Comparison of the measured 11.3~$\mum$ aromatic feature flux (unit:
	    $10^{-13}$ erg~s$^{-1}$cm$^{-2}$) in \cite{Shi2014} from the mid-IR
	    spectra ($f_\text{\rm PAH, spec}$) and  that from the conversion of
	    the fitted \cite{Rieke2009} templates ($f_\text{\rm PAH, SED}$) in
	    \cite{Lyu2017}. 
	}
	\label{fig:pah_str_comp}
    \end{center}
\end{figure}

Figure~\ref{fig:pah_str_comp} compares the 11.3 aromatic feature flux converted
from the fitted \cite{Rieke2009} templates in \cite{Lyu2017} with the values
measured from {\it Spitzer/IRS} spectra by \cite{Shi2014}.  Considering the
measurement uncertainties, the results are consistent within 0.3 dex for most
objects and show a strong correlation. There is only one case above the
correlation that might be a candidate for an additional contribution through
excitation by the nucleus. The consistent results between the SED model
predications and the mid-IR spectral measurements indicate that the excitation
of the 11.3~$\mum$ aromatic feature is dominated by star formation in most of
these systems.

\subsection{Testing Alternative AGN ``Intrinsic'' IR Templates}

Given the consistent SFRs estimated from the SED models with the values
measured by the 11.3~$\mum$ aromatic bands (Figure 9 in \citealt{Lyu2017}) and
the successful reproduction of the host galaxy mid-IR to far-IR properties by
the \cite{Rieke2009} templates shown above, the argument that the intrinsic AGN
IR emission of most quasars can be represented by the SF-corrected
\cite{Elvis1994} template is also validated.  Here we discuss whether the other
versions of AGN ``intrinsic'' IR templates, e.g., as proposed by
\cite{Netzer2007}, \cite{Kirkpatrick2015}, and \cite{Symeonidis2016}, are good
alternative choices to the SF-corrected \cite{Elvis1994} template.

\subsubsection{An Illustration of the Importance of SFG Template Selection}\label{sec:sym-temp}

Recently, \cite{Symeonidis2016} derived an SF-corrected AGN template based on a
sample of 47, radio-quiet PG quasars at $z<0.18$. Even compared with the
original \cite{Elvis1994} quasar template, the \cite{Symeonidis2016} template
has much stronger far-IR emission. However, they used host galaxy luminosities
derived from \citet{Shi2007}, which were based on the \cite{Dale2002} templates
because replacements based on {\it Spitzer} data were not yet available.
Recalling our comparisons of the \cite{Dale2002} and the \cite{Rieke2009} SFG
templates in Section~\ref{sec:SFG-temp}, there is a risk that the cool SED
component they found in the quasars is a result of underestimating the
contribution of star formation to the quasar composite far-IR SED.

\begin{figure}[htp]
    \begin{center}
	\includegraphics[width=1.0\hsize]{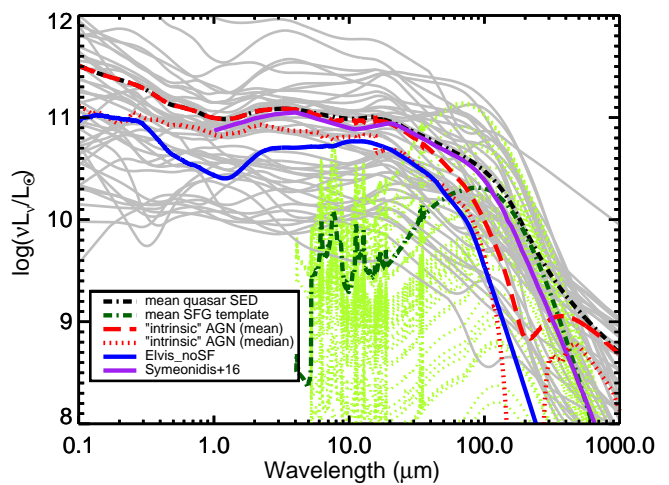} 
	\caption{
	    SEDs for the \cite{Symeonidis2016} PG quasar sample (gray solid
	    curves) and their mean (thick black dashed--dotted curve), as well as
	    the relatively scaled \cite{Rieke2009} SFG templates (light green
	    dotted curves) and their mean (dark green dashed--dotted curve). We
	    derive two ``intrinsic'' AGN SEDs, either by subtracting the mean
	    SFG template from the mean quasar template (mean AGN SED; thick red
	    dashed line) or by subtracting the median SFG template from the
	    median quasar template (median AGN SED; thick red dotted line). The
	    \cite{Symeonidis2016} cooler ``intrinsic'' AGN template and the
	    SF-corrected \cite{Elvis1994} template (normalized at 20~$\mum$ to
	    match the median AGN SED) are also plotted.
	    }
	\label{fig:symeon_check}
    \end{center}
\end{figure}

To test this possibility with the data described in \cite{Lyu2017}, we made a
similar derivation of the ``intrinsic'' AGN IR template based on the same
sample as \cite{Symeonidis2016}.  To derive the quasar continuum, we
interpolated the UV-to-IR photometry logarithmically and smoothed the SED with
a $\Delta\log(\nu)=0.2$ boxcar.\footnote{We note that
\cite{Symeonidis2016} fitted the quasar IR SEDs at $\lambda>22~\mum$ with a
greybody for the far-IR and a power law for the mid-IR (see their Section
3.1). For some quasars with strong host far-IR emission, a local SED minimal
can be seen between $\sim20~\mum$ and $\sim100~\mum$ (see the SEDs of, e.g., PG
0052+251, PG 0844+349, PG 1114+445, and PG 1416$-$129 in their Figure A3). In fact, this
feature is common for many quasars but not always obvious due to the poorly
constrained SEDs at these wavelengths (see our SED decompositions in Figure 5
of \citealt{Lyu2017}). The SED model used by \cite{Symeonidis2016} cannot
reproduce such features due to the smoothly declining nature of their adopted
function, leading to possible overestimations of the far-IR emission of many
quasars.}  With the \cite{Rieke2009} template that gave the closest 11.3~$\mum$
aromatic luminosity as measured by \cite{Shi2007}, we derived a host galaxy
luminosity with the matched and scaled SFG template for each quasar. This host
galaxy IR luminosity was then compared to the observed quasar IR luminosity to
derive a relative scaling of the selected \cite{Rieke2009} template, as shown
in Figure~\ref{fig:symeon_check}. For quasars with only upper limits to the
aromatic flux, we scaled the matched SFG template by adopting one-half of the IR
host template luminosity corresponding to the upper limit,  as in
\cite{Symeonidis2016}.  Then the composite mean SEDs of the quasar sample and
the relatively scaled \cite{Rieke2009} templates were computed. By subtracting
the mean host galaxy SED from the observed mean quasar SED, we derived a final
version of the ``intrinsic'' AGN SED template but with the \cite{Rieke2009} SFG
templates to represent the host galaxy emission. As shown in
Figure~\ref{fig:symeon_check},  this newly derived ``intrinsic'' AGN template,
which is based on the \cite{Symeonidis2016} sample but with the
\cite{Rieke2009} SFG library to convert the 11.3~$\mum$ aromatic strengths to
the host IR luminosities, presents much weaker far-IR emission compared with
the version from \cite{Symeonidis2016}. We also derived another version of the
``intrinsic'' AGN SED template by subtracting the composite median host galaxy
SED from the median quasar SED. Normalized at 20~$\mum$, its far-IR part
matches the SF-corrected \cite{Elvis1994} template reasonably well up to
100~$\mum$.  Beyond 100~$\mum$, our mean (or median) SFG SED has similar
emission strength with the mean (or median) quasar SED, suggesting that the AGN
contribution in the composite quasar emission is weak. A more realistic
replacement of these AGN templates at $\lambda>100~\mum$ is a Rayleigh-Jeans
tail with emissivity proportional to $\lambda^{-1.5}$, which would finally
yield almost the same SED shape at 20--1000~$\mum$ as the SF-corrected
\cite{Elvis1994} template.

Given this result, we suggest that the cooler ``intrinsic'' AGN IR SED derived by
\cite{Symeonidis2016} is at the least a very uncertain conclusion and that a
similar derivation using more appropriate star forming galaxy templates implies
that this characteristic is not common. Consequently, it casts doubt on the
argument of her following work \citep{Symeonidis2017} that the AGN-heated
dust emission would overwhelmingly contribute the far-IR emission of the most
luminous quasars. Additionally, we further confirmed the validity of this
SF-corrected \cite{Elvis1994} template by successfully reproducing a similar
one with a different approach from \cite{Xu2015a}.

\subsubsection{Tests Based on the Aromatic Band Behavior}

As shown in \cite{Lyu2017}, there is a range of intrinsic infrared SEDs for
quasars.  We limit this discussion to the confirmed 52 ``normal" PG quasars as
defined in that paper.

\begin{figure}[htp]
    \begin{center}
	\includegraphics[width=1.0\hsize]{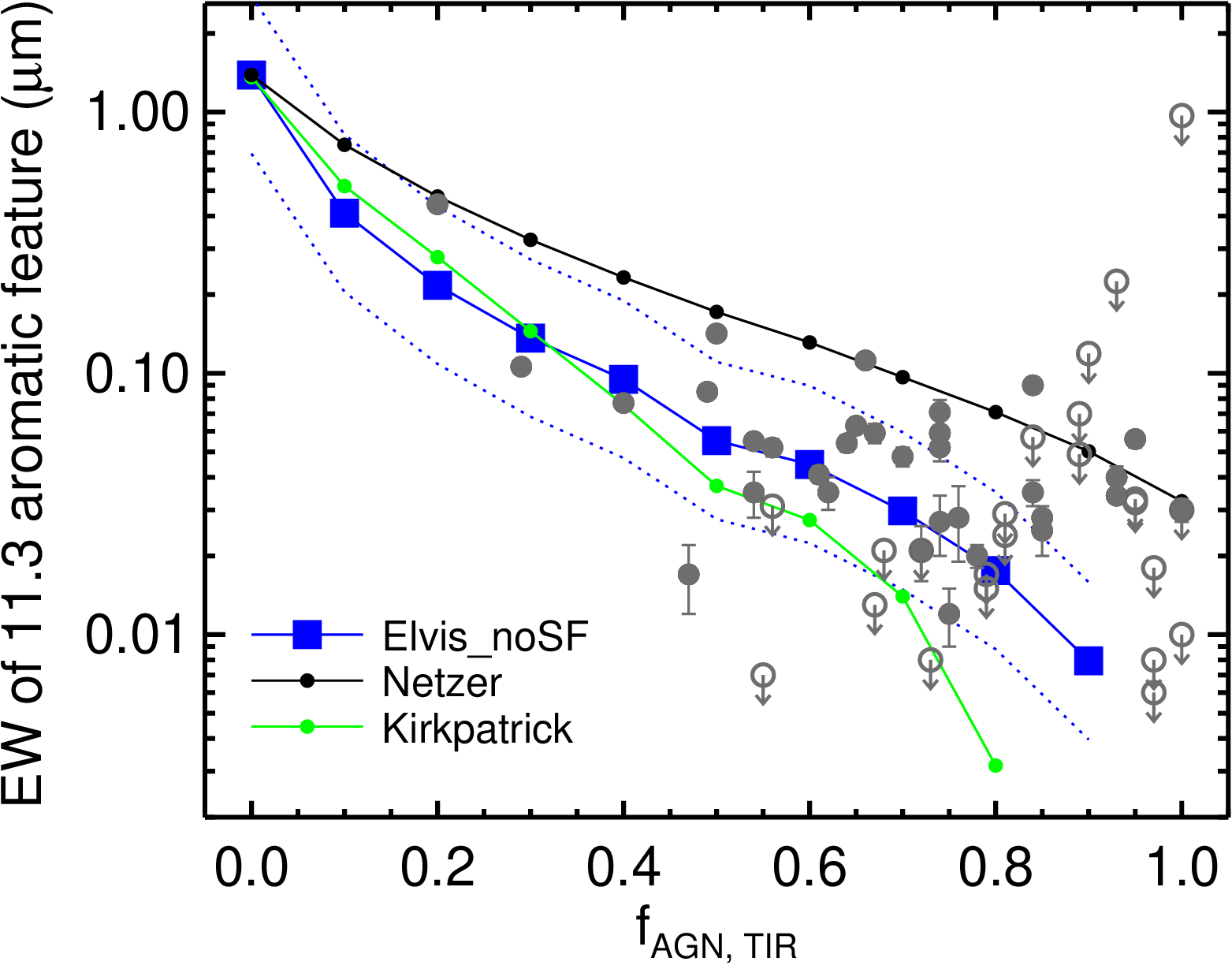}
	\caption{
	    Equivalent width of the 11.3$~\mum$ feature as a function of
	    AGN contribution to the total infrared luminosity, $f_{\rm
	    AGN, TIR}$ for the 57 normal quasars in the PG sample (black filled
	    dots for objects with aromatic feature detections, red open
	    circles for upper limits). We also show the simulated curves by
	    combining the \cite{Rieke2009} $\log(L_{\rm IR}/L_{\odot}))=11.25$
	    star-forming galaxy template with various AGN templates.
	}
	\label{fig:pah_hostf}
    \end{center}
\end{figure}

Since the EW reflects the relative strength of the emission
feature, how the EW of the 11.3~$\mum$ aromatic feature changes with the AGN
contribution to the quasar IR emission budget would provide a method to
distinguish different AGN templates.  In Figure~\ref{fig:pah_hostf}, we compare
the relations between the EW of the 11.3$~\mum$ feature and the
relative AGN contribution in the total infrared emission derived from SED
decomposition of the normal quasar sample and the relations derived from
combining the AGN templates with the \cite{Rieke2009} $\log(L_{\rm
IR}/L_{\odot})$=11.25 SFG template. The observation matches the SF-corrected
\cite{Elvis1994} template reasonably well except for the cases where $f_{\rm
AGN, TIR}>0.8$. In these cases, the AGN contribution is so dominant that the
measurement of the aromatic features becomes difficult, so we are unsure if
this discrepancy is real.  In addition, many quasars with $f_{\rm AGN,
TIR}>0.8$ have host galaxy IR luminosities $\sim10^{10}~L_\odot$, in which case
the $\log(L_{\rm IR}/L_{\odot})$=11.25 SFG template adopted here would
underestimate the EW of the aromatic features (see
Figure~\ref{fig:SFG_temp_comp}).

As shown in Figure~\ref{fig:pah_hostf}, combining the \cite{Kirkpatrick2015}
AGN template with the SFG template to fit the far-IR underpredicts the
11.3$~\mum$ aromatic feature EWs, which supports the idea of that the
\citep{Kirkpatrick2015} templates have substantial host galaxy contamination in
the IR.  In comparison, using the \cite{Netzer2007} AGN template overpredicts
the aromatic feature EWs, suggesting that the host galaxy is over-subtracted
from their quasar average SEDs. As a result, we suggest the assumption in
\cite{Netzer2007} that 50--100$~\mum$ emission of quasars is entirely due to
star formation is too aggressive. Consequently, the SF-corrected
\cite{Elvis1994} template is preferred over these two alternatives.

\section{Discussion}\label{sec:discuss}

\subsection{Does the normal quasar IR template apply to other populations of
type-1 AGNs?}\label{sec:other-agn}

A natural question is whether most, if not all, AGNs have similar intrinsic IR
SEDs. As shown by \citet{Lyu2017}, among unobscured type-1 quasars, some 10\%
or more have a deficiency of the AGN intrinsic infrared emission, making their
SEDs differ significantly from the normal cases. These dust-deficient quasars
can be further grouped into the hot-dust-deficient (HDD) population and the
warm-dust-deficient (WDD) population, possibly connected with different AGN
properties.  In addition, it is established that the AGN SED, including the IR
part, changes from the most luminous quasars to less luminous AGNs in nearby
galaxies \citep{Ho1999, Ho2008, Prieto2010}. Hence the idea of one single IR
template that applies for all kinds of AGNs, even just for quasars, is not
correct.

\begin{figure}[htp]
    \begin{center}
	\includegraphics[width=1.0\hsize]{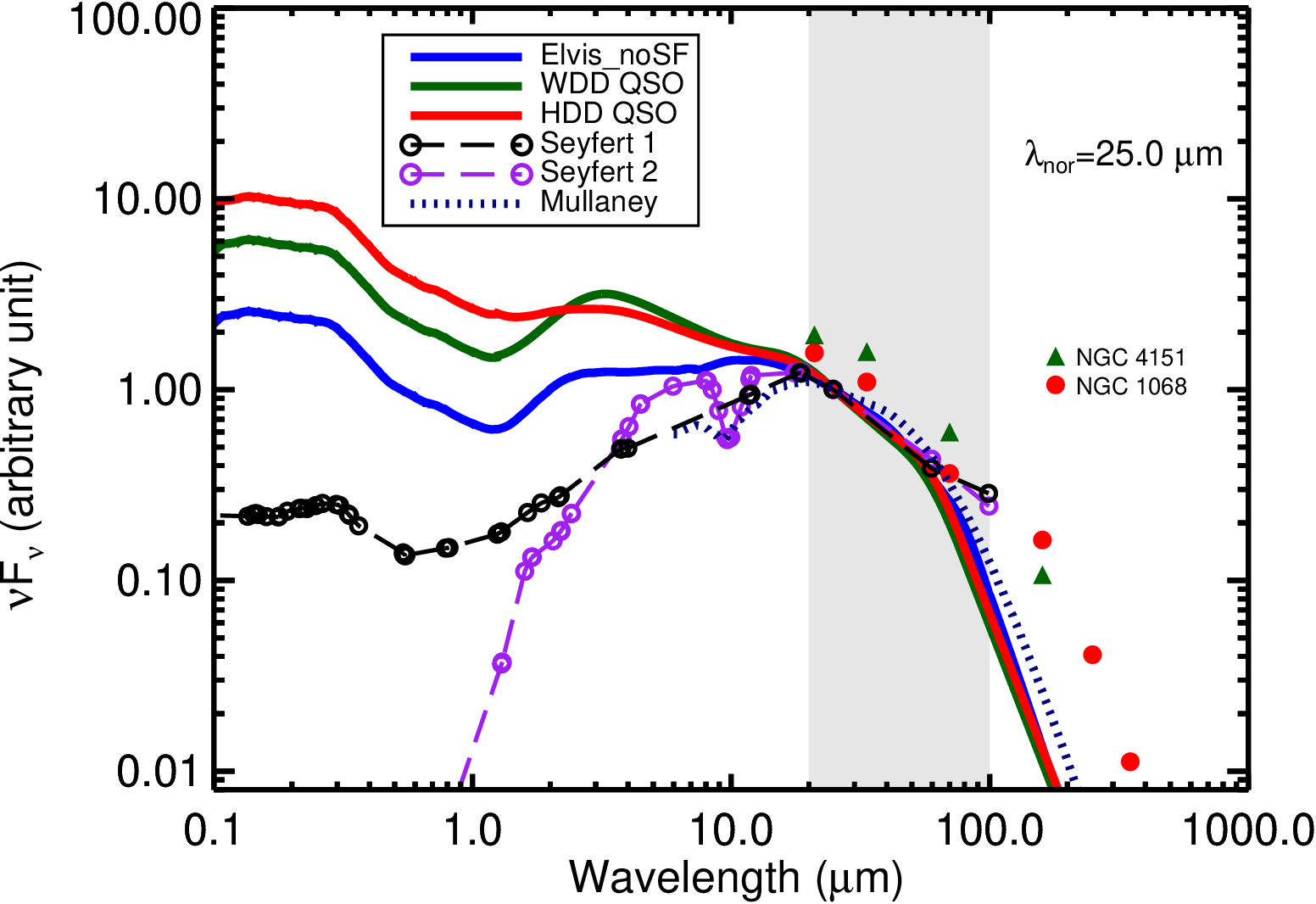} 
	\caption{ 
	Comparison of the AGN templates for quasars and Seyfert galaxies.  We
	highlight the far-IR 20--100$~\mum$ spectral region with a pink
	background.  The normal quasar template is taken from \cite{Xu2015a},
	who removed the far-IR host galaxy contribution from the
	\cite{Elvis1994} template; the hot-dust-deficient (HDD) and
	warm-dust-deficient (WDD) quasar templates are from \cite{Lyu2017}.
	The Seyfert 1 and Seyfert 2 AGN templates are taken from
	\cite{Prieto2010} (we only plot these templates at $\lesssim100~\mum$,
	where there are photometry constraints). The intrinsic IR template for
	relatively low-luminosity AGNs from \cite{Mullaney2011} is also
	presented. We also show the small-beam SEDs of  NGC 4151 and
	NGC 1068.
	    }
	\label{fig:AGN-empirical_SED}
    \end{center}
\end{figure}

Surprisingly, despite the variations in the near- and mid-IR, the intrinsic
SEDs of many different types of AGNs are roughly similar in the far-IR.  In
Figure~\ref{fig:AGN-empirical_SED}, we compare the empirical templates for
normal quasars and dust-deficient quasars \citep{Lyu2017}, the sub-arcsec
resolution average templates derived for nearby Seyfert 1 and Seyfert 2
galaxies \citep{Prieto2010}, the IR intrinsic template of moderate-luminosity
AGNs derived by \cite{Mullaney2011}, and the observed SEDs with the smallest
beams\footnote{\cite{Rieke1972, Rieke1975b, Rieke1975a} presented the smallest
photometric beam measurements of  NGC 1068 and NGC 4151 at
wavelengths longer than $20~\mum$ (except for submillimeter and radio) so far
at 21 and 34$~\mum$ (with a beam size $\sim$6\arcsec). For NGC 1068, we
supplement the 14.3\arcsec aperture {\it Herschel} photometry at 70, 160, 250,
350, and 500~$\mum$ from \citep{Garcia-Gonzalez2016} with a scaling factor of 5.7
to reduce the aperture effects with ground-based data. For NGC 4151, we plot
the 10.2\arcsec aperture photometry at 70 and 160~$\mum$ from the same paper
with a scaling factor of 1.44. } of the archetypal Seyfert galaxies: NGC 1068
(type 2) and NGC 4151 (type 1). Firstly, all three quasar templates have
surprisingly similar 20--100~$\mum$ SEDs with mean deviations of less than 0.07
dex. For Seyfert galaxies, due to their low AGN luminosities, the host galaxy
dust emission still contaminates the far-IR AGN emission. However, all of the
Seyfert AGN templates and SEDs show a decreasing trend of the far-IR SED right
after 20 microns, similar to the behavior of the quasars. In fact, based on
subtraction of host galaxy contamination from the mid-IR continua of Seyfert
galaxies, \cite{Deo2009} also reported a similar turn-over at$\sim20~\mum$.
\cite{Mullaney2011} derived their intrinsic AGN IR template by fitting both
mid-IR spectral data and far-IR photometry of 11 nearby moderate-luminosity
AGNs. We find it shares a similar far-IR SED as the quasar templates. For the
Seyfert sample in \cite{Prieto2010}, their 11.3 aromatic feature
luminosities were found to be $\sim10^{8-9}~L_\odot$ \citep{Sturm2000,Hernan-Caballero2011},
corresponding to the host IR luminosities $\gtrsim10^{11}~L_\odot$. After
subtracting a host galaxy template from the \cite{Prieto2010} Seyfert 1 and
Seyfert 2 AGN templates in the far-IR, their $\gtrsim$70~$\mum$ SED can easily
match that of \cite{Mullaney2011}. As argued at the end of the next
subsection, such an SED similarity in the far-IR might be inherently expected
from simple physics of dust grain emission in the optically thin limit.

\subsection{Implications for the Obscuring Structure}

With the complete characterization of the intrinsic infrared SED templates for
quasars \citep{Xu2015a, Lyu2017} and the further confirmation as presented
in this work, we can analyze the properties of the major dust components that
determine the SED shape. Dust will sublimate at a temperature $T\sim$1800 K in
the innermost regions of the torus.  Thus we use a blackbody with a fixed
temperature at 1800 K to represent this hottest dust component.  Additionally,
we add three dust components to represent the hot, warm, and cold dust
contributions; the first two components are assumed to be black bodies and the
cold dust component is assumed to be a modified blackbody with emissivity
$\beta=1.5$.  Unlike the hottest dust component, the temperatures of the other
three components can be varied freely.  The relative fractions of all the dust
components are free parameters. The emission from the accretion disk is
represented with a power-law component, which can be described as $f_\nu
\propto \nu^{0.3}$ with a break to the $f_\nu \propto \nu^2$ Rayleigh-Jeans
slope at 3$~\mum$ \citep{Honig2010}. We normalize this component to the
template at 0.51$~\mum$.  After subtracting the contribution of this broken
power-law component, the 1.0--1000~$\mum$ SED template is fitted by the
four-component dust model with parameters determined by minimizing the
$\chi^2$.

\begin{figure}[htp]
    \centering
	\includegraphics[width=1.0\hsize]{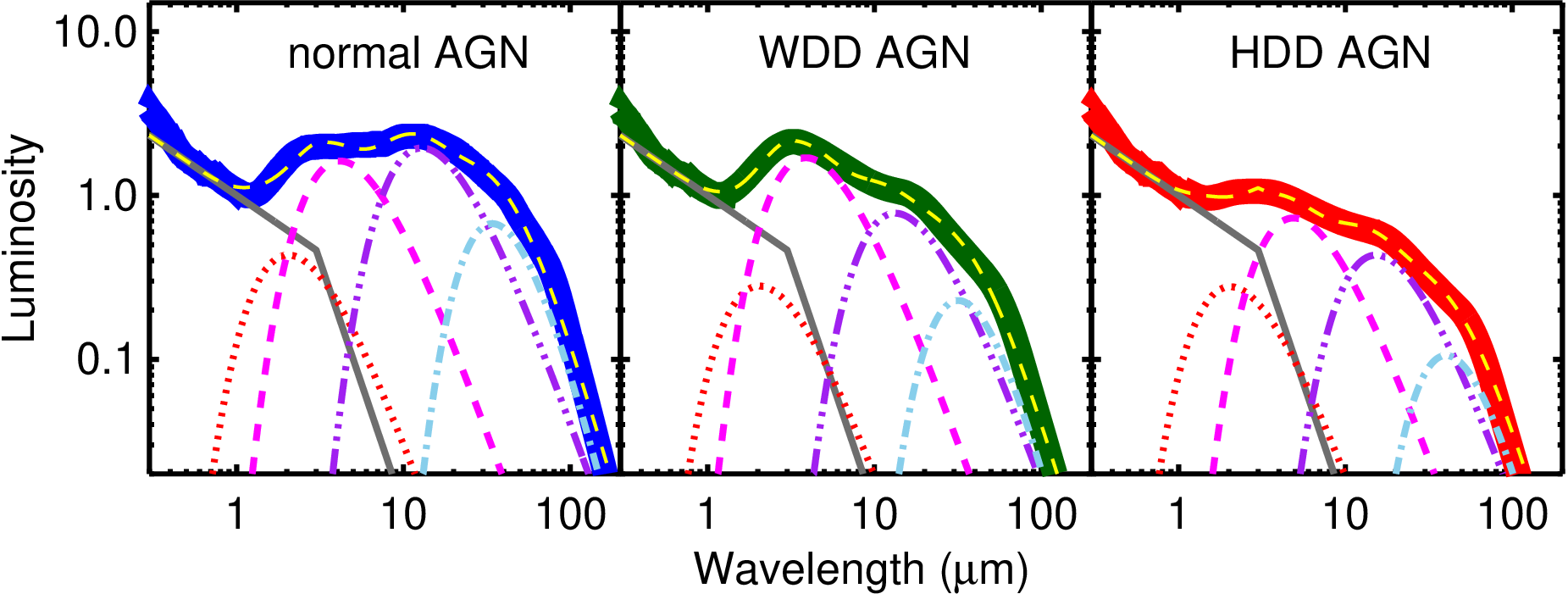} 
	\caption{ 
	  Dust component decomposition of the three empirical AGN templates
	  (normal AGN: blue thick line; WDD AGN: green thick line; HDD AGN: red
	  thick line).  Each template is decomposed into a UV--optical broken
	  power-law component (gray solid lines), a hottest dust component (red
	  dotted lines), a hot dust component (magenta dotted line), a warm
	  dust component (purple dashed--dotted--dotted--dotted lines), and a cold dust
	  component (light blue dashed--dotted lines). The final model templates
	  (yellow dashed lines) are shown against the corresponding empirical
	  templates in each panel.
	}
	\label{fig:dust_decomp}
\end{figure}

The template decomposition results are summarized in
Figure~\ref{fig:dust_decomp} and Table~\ref{tab:dust_decomp}.  Our simple dust
model fits reasonably well and confirms the deficiency of hot or warm dust
emission in corresponding groups of AGNs. In Table~\ref{tab:dust_decomp}, we
quantify the emission strength of each dust component by its total luminosity,
$L_{\rm dust}$, normalized by the emission coming from the accretion disk,
$L_\text{accr.  disk}=L_{\rm AGN}\text{[1 keV--1.3~\mum]}$. The WDD AGNs and
normal AGNs have similar hot dust emission strength with $L_{\rm
dust}/L_\text{accr.  disk}=0.19-0.21$ but the warm dust emission strength of
the former is only about half of the latter.  The HDD AGNs have weaker hot dust
emission ($\sim50\%$) and warm dust emission ($\sim26\%$) compared with the
normal AGNs.  While the relative contributions differ in these AGN templates,
the temperatures of each dust component have narrow ranges with the hot dust
component at $\sim$700--1000~K, the warm dust component at $\sim$200--300~K and
the cold dust component at $\sim$60--80~K. This result suggests that we are
seeing the infrared emission of a number of components of these AGNs that are
nearly always present but in modestly different amounts. 

\capstartfalse
\begin{deluxetable}{lcc}
    \tablewidth{1.0\hsize}
    \tablecolumns{3}
    \tablecaption{Dust Components for the Quasar Intrinsic AGN IR Emission\label{tab:dust_decomp}
    }
    \tablehead{
	\colhead{Dust Component} & \colhead{$T$}  & \colhead{$f_{\rm R}$} \\
	\colhead{(1)} & \colhead{(2)} & \colhead{(3)}     
    }
    \startdata
    \multicolumn{3}{c}{Normal AGN template} \\
    Sublimating dust &    1800 K    &     0.05$\pm$0.01 \\
    Hot dust     & 883$\pm$49 K &     0.19$\pm$0.01 \\
    Warm dust    & 285$\pm$12 K &     0.23$\pm$0.01 \\
    Cold dust    & 77$\pm$5 K   &     0.06$\pm$0.01 \\
    All dust     &       --     &     0.53           \\
    \hline \\[-1.8ex]
    \multicolumn{3}{c}{WDD AGN template} \\ 
    Sublimating dust &     1800 K   &     0.03$\pm$0.02 \\
    Hot dust     & 944$\pm$50 K &     0.21$\pm$0.01 \\
    Warm dust    & 276$\pm$31 K &     0.10$\pm$0.01 \\
    Cold dust    &  83$\pm$13 K &     0.02$\pm$0.01 \\
    All dust     &       --     &     0.36          \\       
    \hline \\[-1.8ex]
    \multicolumn{3}{c}{HDD AGN template} \\ 
    Subliminating dust &     1800 K   &     0.03$\pm$0.01 \\
    Hot  dust    & 752$\pm$77 K &     0.10$\pm$0.01  \\
    Warm dust    & 240$\pm$46 K &     0.06$\pm$0.01  \\
    Cold dust    &  66$\pm$30 K &     0.01$\pm$0.01  \\
    All dust     &       --     &     0.20           
    \enddata
    \tablecomments{
	Column (1): the hottest dust component corresponds the dust close to the
	sublimation distance; Column (2): dust temperature from the decomposition
	model; Column  (3): IR-processed light fraction of each dust component.
	$f_{\rm R}=L_{\rm dust}/L_\text{accr. disk}$. We caculate the 1
	keV--1.25~$\mum$ luminosity of the AGN template as $L_\text{accr. disk}$
	and the 1.25--1000~$\mum$ luminosity as $L_\text{dust}$.
    }
\end{deluxetable}
\capstarttrue

Given the decomposition results, we can also make order-of-magnitude
estimations of the physical sizes of these dust components. Assuming blackbody
emission with temperature $T_{\rm d}$, the dust emission luminosity $L_{\rm d}$
can be approximated by the Stefan--Boltzmann law. Introducing a dust-covering
factor $f_{\rm c}$ and a characteristic physical scale $r_{\rm d}$, then we
have
\begin{equation}
    L_{\rm d} \sim 4\pi r_{\rm d}^2 ~f_{\rm c}  ~\sigma T_{\rm d}^4 ~~~,
\end{equation}
where $\sigma$ is the Stefan--Boltzmann constant. The dust-covering factor of
each component can be assumed to be $\lesssim1$. Substituting typical dust
temperatures and the IR-processed light fractions of various dust components,
for an AGN with luminosity $L_{\rm AGN} = 10^{11}~L_\odot$, the characteristic
physical scales of the hot, warm, and cold dust emission are of the order of
0.01--0.1, 1, and 10~pc. All these values scale with the square-root of AGN
Luminosity ($r_{\rm d} \propto L_{\rm AGN}^{0.5}$).  The typical physical size
of the hot dust emission estimated from the SED analysis is consistent with the
estimation from dust sublimation \citep[e.g.,][]{Laor1993}, suggesting that the
subliming dust and hot dust may be part of a continuous distribution.  The
relatively weak far-IR emission of quasars suggests a compact torus at
sub-kiloparsec scales, which was originally suggested by \cite{Pier1992,
Pier1993} from a theoretical analysis. Only for the most luminous quasars
($L_{\rm AGN}\gtrsim 10^{14}~L_\odot$), the dust far-IR emission heated by the
AGN could extend into kiloparsec scales, but maybe only marginally so if the
relatively decreasing mid-IR and far-IR emission of the AGN is common in these
systems (see Section 6.3 of \citealt{Lyu2017}). 

The results from the simple SED analysis above show that the AGN-heated dusty
structures have a wide temperature distribution with diverse
physical scales. Although the AGN obscuration structures are often pictured
as a doughnut-like ring (or a torus), we are actually unsure of the size
and structure of their outer part. In the mid-IR, high-resolution observations
support a compact geometry of the mid-IR emission region for local AGNs
(see, e.g., \citealt{Asmus2014} and references therein).
However, as suggested by, e.g., \cite{Antonucci2015}, one should be cautious
when referring to a size of the torus since it depends on the observed wavelength.
The Atacama Large Millimeter Array (ALMA) has the resolution to possibly image
the torus structures in nearby systems \citep{Garcia-Burillo2016,
Imanishi2016}. Nevertheless, whether these dusty molecular clouds are obscuring
the nuclei and heated up in the submillimeter bands by absorbing AGN emission
is not easy to tell. The similar intrinsic far-IR SEDs of AGNs discussed in
this work provide some observational constraints on this topic.

In the far-IR, the dust emission is expected to be in the optically thin limit.
As suggested by \cite{Ivezic1997}, the SED shape under this condition is not
sensitive to the geometry but the temperature of the dust grains (see their
Section 5). If similar grain properties can be assumed, the identical
equilibrium temperatures of optically thin dust, which would result in similar
intrinsic AGN far-IR SEDs, are naturally expected once the spectral shape of
the incoming radiation is settled \citep[e.g.,][]{Laor1993}. In fact, the
incoming light that is absorbed by these dust grains in quasars and Seyfert
nuclei share comparable SEDs (dominated by the emission of, e.g., the thin-disk
accretion; \citealt{Yuan2014}).  Meanwhile, the geometric structures of the
torus, whose outer part could be potentially mixed with the galactic ISM, are
allowed to be somewhat diverse.

\section{Summary}\label{sec:summary}

The discrepancies in determinations of the intrinsic far-IR SED of luminous
AGNs have been evaluated in this paper. We found that the SF-corrected
\cite{Elvis1994}-like AGN template is the most-likely correct selection for
most type-1 quasars. This conclusion is supported by the following evidence.
\begin{enumerate}
    \item {\it Energy balance}. We assumed the IR (1.3--1000~$\mum$) emission of
	the AGN comes from the dust-reprocessed black hole accretion emission
	at 0.0012 (1 keV)--1.3~$\mum$ and the IR-processed light fraction,
	$f_{\rm R}$, should be consistent with the observed dust-covering
	factor $\lesssim0.65$ for luminous quasars. Adopting the \cite{Elvis1994}
	template to represent the SED of the accretion disk emission with a
	proper scaling, we found only a small number of empirical IR templates
	yielded a matched $f_{\rm R}$ (e.g., 0.53-0.55 for \citealt{Elvis1994}
	as well as its SF-corrected version, 0.56 for \citealt{Symeonidis2016},
	0.46 for \citealt{Netzer2007}). In comparison, the
	AGN templates proposed by \cite{Polletta2007}, \cite{Hanish2013}, and
	\cite{Kirkpatrick2015} have $f_{\rm R}\gtrsim0.7$, suggesting that
	they are unlikely valid for luminous type-1 objects in general.  
    
    \item {\it SED decomposition of the PG quasars}.
	Based on the results of \cite{Lyu2017}, we found that (1) the far-IR
	SEDs of PG quasars are fitted well by the SF-corrected \cite{Elvis1994}
	AGN template combined with a \cite{Rieke2009} SFG template; (2) the
	best-matched SFG templates preserve the luminosity-dependent SED shapes
	as seen in IR-luminous star-forming galaxies; (3) the predicted
	11.3~$\mum$ aromatic strengths from the SFG templates match the
	measurements from the quasar mid-IR spectra; (4) the observed relation
	between the EWs of the 11.3~$\mum$ aromatic feature and
	the AGN contributions to the total quasar infrared luminosities is also
	matched by the mock SED simulation by the SF-corrected \cite{Elvis1994}
	and the \cite{Rieke2009} templates.
\end{enumerate}

Compared with the observations, the mock composite SEDs with the
\cite{Netzer2007} AGN template overestimate the EW of the
11.3~$\mum$ aromatic feature, suggesting the assumption that the far-IR
emission of quasars is totally SF-dominated is not completely true. 
Meanwhile, the cooler IR emission nature of the \cite{Symeonidis2016}
intrinsic AGN template is a result of the questionable adoption of the
\cite{Dale2002} galaxy templates to relate the strengths of the aromatic
feature to the galaxy far-IR emission for (U)LIRGs. The correct conversion
between the aromatic flux and the galaxy far-infrared luminosity, plus
high-quality mid-IR spectra, are required to avoid an underestimation of host
galaxy contribution to the far-IR emission of quasars.

There is no one single IR template that can be applied to all kinds of AGNs,
even just for type-1 quasars. Nevertheless, normal quasars, dust-deficient
quasars and Seyfert nuclei have similar intrinsic AGN far-IR SED shapes at
$\lambda>20~\mum$, which may indicate a similar emitting character in the outer
part of the AGN-heated dusty structures. In fact, the similar intrinsic
AGN far-IR emission SEDs for these objects are naturally expected when the
responsible dust grains are in the optically thin limit.

Based on the decomposition of the intrinsic AGN IR templates for the type-1
quasar population, we found that four dust components with similar temperatures
($T\sim$1800, 700--1000, 200--300, and 60--80 K) can explain the diversity of the
intrinsic AGN IR emission properties by changing their relative contributions.
The weak emission of the AGN-heated cold dust component suggests a compact
torus at sub-kiloparsec scales in the far-IR for most quasars.

The intrinsic AGN templates for normal quasars \citep{Xu2015a} and
dust-deficient quasars \citep{Lyu2017} are provided in the appendix.

\acknowledgments

This work was supported by NASA grants NNX13AD82G and 1255094. We thank Allison
Kirkpatrick, Alexandra Pope, Anna Sajina, Richard Green, and Aigen Li for
helpful discussions and/or comments on an early draft of this paper, and Yong
Shi for sharing the measurements of the {\it Spitzer}/IRS spectra of the PG
sample in \cite{Shi2014}. We also appreciate the comments from the referee that
have helped us to improve the clarity of our writing.

\appendix
\section{The AGN intrinsic SED templates}\label{app:temp}

\cite{Xu2015a} derived the intrinsic AGN template for normal quasars based on
the \cite{Elvis1994} template. The validity of this template has been
demonstrated for the most luminous quasars at $z\gtrsim5$ \citep{Lyu2016},
type-1 AGNs at $z\sim$0.7--2.5 \citep{Xu2015a}, and PG quasars at $z<0.5$
\citep{Lyu2017}. In \cite{Lyu2017}, we have derived the intrinsic AGN templates
for WDD and HDD quasars and shown that these dust-deficient AGNs can be found at
$z\sim$0--6. In Table \ref{tab:lyu-temp}, we provide the 0.1--1000~$\mum$ SEDs of
these three templates.

\capstartfalse
\begin{deluxetable}{ccccc}[!htbp]
    \tablewidth{1.0\hsize}
    \tablecolumns{5}
    \tablecaption{AGN Intrinsic Templates\label{tab:lyu-temp}
    }
    \tablehead{
	\colhead{ \scriptsize $\log(\lambda/\micron)$} & 
	\colhead{ \scriptsize$\log(\nu/\text{Hz)}$}  & 
	\colhead{\scriptsize$\log(\nu F_{\nu,\text{normal}})$}& 
	\colhead{\scriptsize$\log(\nu F_{\nu,\text{HDD}})$}&
	\colhead{\scriptsize$\log(\nu F_{\nu,\text{WDD}})$} 
	\\
	\colhead{(1)} & 
	\colhead{(2)} & 
	\colhead{(3)} & 
	\colhead{(4)} & 
	\colhead{(5)} 
    }
    \startdata
  -1.00  & 15.48  & 0.575  & 0.574  & 0.571 \\
  -0.99  & 15.47  & 0.578  & 0.576  & 0.574 \\
  -0.98  & 15.46  & 0.579  & 0.578  & 0.576 \\
  -0.97  & 15.45  & 0.588  & 0.587  & 0.585 \\
  -0.96  & 15.44  & 0.595  & 0.594  & 0.591 \\
   \dots
    \enddata
    \tablecomments{
	The template luminosities are normalized at 1.25 $\mu$m.  For all of the
	templates at $\lambda>100~\mum$, a modified blackbody with temperature
	at 118 K and dust emissivity 1.5 is scaled to match the observed SED.
	The 0.1--1.25$~\mum$ SEDs of the HDD and WDD templates are assumed to be
	the same as \cite{Elvis1994}.\\ (This table is available in its
	entirety in machine-readable form. A portion is shown here for guidance
	regarding its form and content.)
    }
\end{deluxetable}
\capstarttrue

\section{Pure star-forming galaxies in the GOALS sample}
\label{app:goals}

The Great Observatories All-sky LIRG Survey (GOALS; \citealt{Armus2009})
provided the community a comprehensive dataset for over 200 (U)LIRGs in the
local Universe. We selected 101 pure star-forming galaxies from this sample by
removing any objects with an X-ray cross-identification according to the
NASA/IPAC Extragalactic Database (NED), identified as AGN in the literature, or
presenting mid-IR \NeV~emission lines. We obtained the {\it Spitzer}/IRS
low-resolution spectra for all these objects from the {\it Combined Atlas of
Sources with Spitzer IRS Spectra} (CASSIS; \citealt{cassis}). If necessary, we
scaled the Short-Low (SL) module spectra to match the Long-Low (LL) module
spectra to make a continuous mid-IR continuum. The final combined spectra were
then analyzed by the IDL program {\it PAHFIT} \citep{Smith2007}. Finally, the
flux of the 11.3 aromatic feature complex was derived for each object. Adopting
the 8--1000~$\mum$ IR luminosities in \cite{Armus2009}, we calculated the
luminosity conversion factor from the 11.3 aromatic feature complex to the
total IR emission for each star-forming galaxy and binned individual
measurements as a function of the IR luminosity. The standard deviations are
used to show the dispersions of the observed conversion factors in
Figure~\ref{fig:SFG_temp_comp}. Our sample of pure SFGs in GOALS as well as
corresponding measurements are presented in Table~\ref{tab:goals}.

\capstartfalse
\begin{deluxetable}{lcccc}[!htbp]
    \tablecolumns{5}
    \tablecaption{Properties of 101 SFGs in the GOALS sample \label{tab:goals}
    }
    \tablehead{
	\colhead{Name} &
	\colhead{$z$}  & 
	\colhead{$\log (L_{\rm PAH}/L_\odot)$} 	& 
	\colhead{$\log (L_{\rm IR}/L_\odot)$}    &
	\colhead{$f_{\rm conv.}$} \\
	\colhead{(1)} & 
	\colhead{(2)} & 
	\colhead{(3)} & 
	\colhead{(4)} & 
	\colhead{(5)} 
    }
    \startdata
    RAS F19297-0406  &   0.0857   &   9.30   &  12.40  &  0.0008\\ 
    IRAS 19542+1110  &   0.0650   &   9.38   &  12.00  &  0.0024\\ 
    IRAS 17132+5313  &   0.0509   &   9.72   &  11.90  &  0.0066\\ 
       CGCG 448-020  &   0.0361   &   9.39   &  11.90  &  0.0031\\ 
      ESO 593-IG008  &   0.0487   &   9.70   &  11.90  &  0.0062\\ 
   \dots
\enddata                                    
    \tablecomments{
	Column (1): the object name; Column (2): redshift from NED; Column (3): the luminosity of the 11.3 aromatic emission
	as measured from the mid-IR spectrum; Column
	(4): the IR luminosity of the object in \cite{Armus2009}; Column (5): the luminosity conversion factor between
	the 11.3 aromatic emission and the total IR emission. $f_{\rm conv.}=L_{\rm PAH}/L_{\rm IR}$.\\
	(This table is available in its entirety in machine-readable form. A
	portion is shown here for guidance regarding its form and content.)
    }
\end{deluxetable}
\capstarttrue

\bibliographystyle{apj.bst}

\end{document}